\documentclass[useAMS,usenatbib]{mn2e}

\usepackage{graphicx}
\usepackage{epsfig}
\usepackage{amssymb,amsmath}
\usepackage{multirow}
\usepackage{mathrsfs}
\usepackage{times}
\usepackage{setspace}
\usepackage[dvipsnames]{xcolor}
\usepackage{soul}

\newcommand{\green}[1]{{#1}}

\usepackage[colorlinks, breaklinks, citecolor=blue]{hyperref}

\title[Metallicity Fluctuation Statistics]{Metallicity Fluctuation Statistics in the Interstellar Medium and Young Stars. I. Variance and Correlation}

\author[Krumholz \& Ting]{Mark R. Krumholz$^1$\thanks{mark.krumholz@anu.edu.au}
and Yuan-Sen Ting$^{1, 2, 3, 4}$\thanks{yuan-sen.ting@anu.edu.au}
\\ \\
$^1$ Research School of Astronomy \& Astrophysics, Australian National University, Canberra, ACT, Australia\\
$^2$ Institute for Advanced Study, Princeton, NJ 08540, USA\\
$^3$ Department of Astrophysical Sciences, Princeton University, Princeton, NJ 08544, USA\\
$^4$ Observatories of the Carnegie Institution of Washington, 813 Santa Barbara Street, Pasadena, CA 91101, USA
}

\begin{document}
\maketitle
\label{firstpage}
\begin{abstract}
The distributions of a galaxy's gas and stars in chemical space encodes a tremendous amount of information about that galaxy's physical properties and assembly history. However, present methods for extracting information from chemical distributions are based either on coarse averages measured over galactic scales (e.g., metallicity gradients) or on searching for clusters in chemical space that can be identified with individual star clusters or gas clouds on $\sim 1$ pc scales. These approaches discard most of the information, because in galaxies gas and young stars are observed to be distributed fractally, with correlations on all scales, and the same is likely to be true of metals. In this paper we introduce a first theoretical model, based on stochastically-forced diffusion, capable of predicting the multi-scale statistics of metal fields. We derive the variance, correlation function, and power spectrum of the metal distribution from first principles, and determine how these quantities depend on elements' astrophysical origin sites and on the large-scale properties of galaxies. Among other results, we explain for the first time why the typical abundance scatter observed in the interstellar media of nearby galaxies is $\approx 0.1$ dex, and we predict that this scatter will be correlated on spatial scales of $\sim 0.5-1$ kpc, and over time scales of $\sim 100-300$ Myr. We discuss the implications of our results for future chemical tagging studies.
\vspace{0.1in}
\end{abstract}

\begin{keywords}
diffusion --- Galaxy: abundances --- galaxies: abundances --- galaxies: ISM --- ISM: abundances --- stars: abundances
\end{keywords}


\section{Introduction}

The chemical content of galaxies, in both the stellar and gaseous phases, provides a unique window into the history of their formation and evolution. Almost all elements heavier than hydrogen and helium are manufactured in stars and then ejected into the interstellar medium (ISM) during the final stages of stellar evolution. Once ejected they can be measured \textit{in situ} in the interstellar gas wherever it is illuminated by ionising radiation. Some of these metals will also be incorporated into a next generation of stars, where they can be observed in stellar atmospheres. Since individual atoms are not altered during their time in the ISM or by incorporation into a stellar atmosphere, measurements of their abundances provide a complete record of prior star formation and nucleosynthesis.

These data have begun to accumulate in quantity. On the gaseous side, the advent of massively-multiplexed spectrographs and integral field units, supplemented by large radio surveys within the Milky Way, has made it possible to measure gas phase abundances as a function of position in the ISM of large samples of nearby galaxies \citep[e.g.,][]{croxall09a, sanders12a, sanchez14a, balser15a, bresolin15a, berg15a, ho15a, james16a, vogt17a}. With the aid of lensing, these data are even beginning to become available for the high-redshift universe \citep[e.g.,][]{jones13a, yuan15a, leethochawalit16a}. Traditionally these data have been binned by azimuth in order to constrain the mean radial gradient of metal abundance within a galaxy disc. While this is indeed an important constraint for models of galaxy formation, these new instruments deliver such a large number of pixels or individual ionised regions in a single galaxy that it has become possible to study higher-order statistics, such as the dispersion and correlation of elements with azimuth.

On the stellar side, large spectroscopic surveys of Milky Way stars such as APOGEE \citep{holtzman15a,majewski15a,sdss-collaboration17a}, the Gaia-ESO Public Spectroscopic Survey \citep{gilmore12a}, and GALAH \citep{de-silva15a} are in the process of delivering samples of high-quality abundance measurements for $\sim 10^5-10^6$ stars. Individual open clusters show extremely small abundance variations at the level of $\sim 0.02 - 0.03$ dex \citep[e.g.,][]{bovy16a, liu16a, ness17a}. This homogeneity has led to the idea of chemical tagging: reconstructing the formation history of the Milky Way by identifying stars that formed in the same cluster via their near-identical chemical abundances, but that have since escaped their natal sites and become phase-mixed in the Galactic field \citep{freeman02a, bland-hawthorn10a, bland-hawthorn10b,ting15a}.

In principle the statistics of metallicity distributions in the gas phase, and the clustering of stars in chemical space, provide very strong constraints on the formation history and star formation structure of the galaxy in which they are measured, and on the astrophysical origin sites of the elements. There are a few examples in the literature of attempts to exploit these constraints to identify otherwise-uncertain origin sites, particularly for $r$-process elements \citep[e.g.,][]{matteucci14a, van-de-voort15a, shen15a, hirai17a}, and to measure the Milky Way cluster mass function in the past through the clustering/abundance-correlation of stars \citep{ting16a,ness17a}. For the most part, however, we have limited ability to make use of data on the abundance statistics of gas or young stars, because we lack a theoretical model capable of predicting and interpreting the mapping between chemical and physical space.

For example, suppose that we were to identify two stars on opposite sides of the Milky Way whose abundances differ by $\sim 0.01 - 0.05$ dex across a wide range of elements. What should we then infer about how closely together in space and time those two stars formed? Did they form within 1 pc of one another? 100 pc? 1 kpc? Within 1 Myr, or 100 Myr? The current approach of attempting to identify discrete star clusters \citep[e.g.,][]{bland-hawthorn10b, bland-hawthorn10a, karlsson13a,ting15a,ting16a} is clearly inadequate, since modern observations of star formation reveal that star clusters are not distinct, discrete objects; instead, they simply represent the peaks of a continuous distribution of young stars that is correlated on many scales (e.g., \citealt{gouliermis10a, gouliermis12a, gouliermis17a, grasha17a, grasha17b}; for a review, see \citealt{krumholz14c}). Presumably the metallicity distribution is similarly continuous and correlated across many scales. Similarly, suppose that we measure the metallicity distributions in the ISM of a pair of nearby galaxies, and discover that in one the metallicity scatter is twice as large as the other, and has double the correlation length. What does that tell us about the differences in these galaxies' formation histories?

Existing theoretical models are far from being able to answer these questions. One major approach to chemical modelling has been to develop semi-analytic models in which a galaxy is broken into radial zones, tracking production of elements within zones and (sometimes) transport between them \citep[e.g.,][]{spitoni10a, spitoni11a, matteucci14a, forbes14a, pezzulli16a}. While this approach allows the efficient study of radial gradients, it is obviously unable to make any predictions about higher order statistics. 

Cosmological simulations including chemical enrichment and tracking \citep[e.g.,][]{few12a, pilkington12a, brook12a, minchev13a, van-de-voort15a, shen15a, grand15a, hopkins17a, escala17a} can in principle make such predictions. In practice, however, they are unable to do so due to numerical limitations. \green{For large galaxies like the Milky Way}, these simulations achieve resolutions of $\sim 100$ pc at best\green{, and even in dwarfs the best possible resolution is $\gtrsim 10$ pc.\footnote{\green{When discussing the effective resolutions of cosmological simulations that use adaptive mesh or Lagrangian methods, it is important to distinguish between the peak resolution and the average resolution in galaxy discs. Metallicity statistics are always measured as spatial averages in Eulerian coordinates, so the ability of a simulation to capture metallicity variations depends on the latter, not the former. As an illustration of the importance of this distinction, note that the Milky Way-analog simulations reported by \citet{hopkins17a} have a maximum spatial resolution of $\sim 1$ pc, but a mass resolution of $7000$ $M_\odot$. At the mean number density of the Milky Way's ISM, $\sim 1$ H nucleus cm$^{-3}$, a resolution element of this mass occupies a volume of $(60\mbox{ pc})^3$. Since meaningful statistics can only be measured over a minimum of a few resolution elements, the effective spatial resolution of these simulations for the purposes of measuring metallicity statistics is $\gtrsim 100$ pc, two orders of magnitude larger than the peak resolution.}} This} makes their predictions for metallicity distributions on small scales within a galaxy (as opposed to between galaxies or in galaxy halos) very sensitive to poorly-constrained subgrid prescriptions for unresolved transport processes \citep{revaz16a}. Moreover, they are an order of magnitude too coarse to resolve the natural correlation length of metallicity variations that we derive below.  While some authors have published simulations of isolated discs or portions thereof with enough resolution to address metallicity variation within a galaxy, thus far these have been used to quantify mixing rates, rather than to provide a full exploration of metallicity statistics \citep[e.g.][]{de-avillez02a, yang12a, feng14a, petit15a}; indeed, due to their high resolution, it is not at present feasible to run such simulations over cosmological times, as would be required for this purpose.

The goal of this series of papers is to develop a new approach to the problem of metallicity statistics within the ISM of galaxies and the young stars it forms. Given that numerical simulations are still far from being able to solve this problem, let alone do so enough times to provide large statistical samples, our approach will be primarily analytic and semi-analytic. In this first paper we model the metal distribution in galaxies as the result of a stochastically-forced diffusion process: metals are injected randomly by star formation events, and then transported away from their formation sites and mixed by interstellar turbulence. While this model is obviously an oversimplification of the true ISM, it has the virtue that it is simple enough to admit exact analytic calculation of some of the statistics of greatest interest -- the variance of metallicities and the correlation of metallicity in space and time -- thereby allowing us to determine the relationship between these statistics and the physical properties of galaxies and the nucleosynthetic sites of elements. In the next paper (Ting \& Krumholz, in preparation), we use this framework to conduct semi-analytic numerical calculations that enable us to address element-to-element correlations and their statistics. 

The plan for the remainder of this paper is as follows. In \autoref{sec:stats} we introduce our simple formal model for the metal distribution in a galaxy, and derive its statistical properties. In \autoref{sec:implications} we use the model to deduce the relationship between metallicity statistics and galaxy properties. Finally, we discuss and summarise our conclusions in \autoref{sec:conclusion}.

\section{Statistics of Metal Fields}
\label{sec:stats}

\subsection{Model System}

We consider a simple model for the metallicity distribution in some region of a galaxy. We assume that the distribution is a balance between injection events, which we model as $\delta$ function-like additions of metal, and spreading of metals, which we model as a linear diffusion process. In reality linear diffusion, meaning a diffusion coefficient that is independent of scale, is a significant oversimplification of turbulent transport \citep[e.g.,][]{pan10a, pan11a, yang12a, colbrook17a}. However, since our goal is not precise statistics but instead a first calculation that is accurate enough to yield scalings and approximate values but simple enough to solve analytically, we will ignore this complication. \green{Numerical simulations show that the main difference between true turbulent mixing and linear diffusion is that, in a turbulently-mixed field, rare patches of poorly-mixed gas can persist for much longer than would be expected for diffusive mixing. As a result, the abundance distribution develops significantly non-Gaussian tails. To minimise the error we make by ignoring this effect, we will in this paper focus on statistics such as the metallicity dispersion and two-point correlation function that are mostly sensitive to the central parts of the metallicity distribution, rather than to the extreme tails.
}

For injection, we consider a portion of a galactic disc, and let $\Gamma$ be the constant rate per unit area of injection events, each of which adds a mass $m_X$ of some metal $X$. \green{The assumption of a spatially-uniform injection rate is reasonable for nucleosynthetic sites associated with older stellar populations, but is probably not reasonable for type II supernovae, which are highly clustered; we attempt to account for this effect approximately below, in \autoref{sssec:rate_variance}, and for now we defer further discussion of this point} The value of $m_X$ is drawn from a specified, position- and time-independent distribution $p_m(m_X)$, which we require to have finite expectation value and variance. Injected metals are distributed over an injection kernel $f_{\rm inj}(\mathbf{x},t)$ which has units of inverse area times time, and unit integral. The injection events are uniformly distributed in both space and time. Once injected, the metallicity distribution diffuses with a constant diffusion rate coefficient $\kappa$. The metal surface density $\Sigma_X(\mathbf{x},t)$ as a function of two-dimensional position $\mathbf{x}$ and time $t$ then is fully described by the stochastic partial differential equation (PDE)
\begin{equation}
\frac{\partial}{\partial t} \Sigma_X = \kappa \nabla^2 \Sigma_X + \sum_i m_{X,i} f_{\rm inj}(\mathbf{x}-\mathbf{x}_i, t-t_i),
\end{equation}
where the sum is over the masses $m_{X,i}$, positions $\mathbf{x}_i$, and times $t_i$ of the injection events. The values $m_{X,i}$, $\mathbf{x}_i$, and $t_i$, as well as the number of injection events, are random variables. Within any region of area $A$, the number of injection events over some time interval $T$ is drawn from a Poisson distribution $P_\lambda(N)$ with expectation value $\lambda = \Gamma A T$, while $\mathbf{x}_i$ and $t_i$ are drawn from uniform distributions over the area $A$ and time interval $T$, respectively, and $m_{X,i}$ is drawn from $p_m(m_X)$. We are interested in characterising the statistical properties of the resulting metal field $\Sigma_X$.

As a first step, let us make a change of variables to non-dimensionlise the problem. We define
\begin{eqnarray}
\mathbf{r} & = & \mathbf{x}/x_s \\
\tau & = & t/t_s \\
S_X & = & \Sigma_X / ( \langle m_X \rangle / x_s^2 ),
\end{eqnarray}
where $\langle m_X\rangle$ is the expectation value of the distribution $p_m(m_X)$. We choose our scaling factors $x_s$ and $t_s$ to be
\begin{eqnarray}
\label{eq:xs}
x_s & = & \left(\frac{\kappa}{\Gamma}\right)^{1/4} \\
\label{eq:ts}
t_s & = & \sqrt{\frac{1}{\Gamma\kappa}}.
\end{eqnarray}
With these choices, the evolution equation is
\begin{equation}
\label{eq:stoch_diffusion}
\frac{\partial}{\partial \tau} S_X = \nabla_r^2 S_X + \sum_i w_i s_{\rm inj}(\mathbf{r}-\mathbf{r}_i, \tau-\tau_i),
\end{equation}
where $w_i$ is a random variable drawn from the distribution $p_w(w) \sim p_m(\langle m_X\rangle w)$ (i.e., $p_w(w)$ is simply $p_m(m_X)$ scaled to have unit expectation value), the non-dimensional injection kernel is
\begin{equation}
s_{\rm inj} = f_{\rm inj} x_s^2 t_s,
\end{equation}
and $\nabla_r$ implies differentiation with respect to $\mathbf{r}$ rather than $\mathbf{x}$. With this change of variables, the expectation value $\lambda$ for the number of events in is $\lambda = 1$ per unit area per unit time, when area is measured in units of $x_s^2$ and time in units of $t_s$. We discuss the likely physical values of $x_s$ and $t_s$ in more detail in \autoref{ssec:phys_scales}, where we show that, for Milky Way conditions and for species primarily produced by type II supernovae, $x_s \sim 100$ pc and $t_s \sim 30$ Myr.

\subsection{Formal Solution}

Consider an injection shape function $s_{\rm inj}$ that is a $\delta$ function in time and a Gaussian in space, i.e.,
\begin{equation}
s_{\rm inj}(\mathbf{r}_i,\tau_i) = \frac{1}{2\pi \sigma_{\rm inj}^2} \exp\left[-\frac{|\mathbf{r}-\mathbf{r}_i|^2}{2\sigma_{\rm inj}^2}\right] \delta(\tau - \tau_i).
\end{equation}
\green{While there is no particular reason to assume that the injection distribution is Gaussian, this choice is reasonable because the action of diffusion is such that even a highly non-Gaussian injection profile will result in a nearly-Gaussian distribution at spatial scales much larger than the initial injection region. The advantage of choosing a Gaussian is that it makes the solution} particularly simple because the action of the diffusion operator on a Gaussian profile is simply to increase its dispersion, and the operator is linear. In this case one may verify by direct substitution that the solution is
\begin{equation}
\label{eq:formal_solution}
S_X(\mathbf{r},\tau) = \sum_{i, \tau_i<\tau} w_i \phi(|\mathbf{r}-\mathbf{r}_i|, \tau-\tau_i),
\end{equation}
where
\begin{equation}
\phi(r,\tau) \equiv \frac{1}{4\pi(\tau+\tau_0)} \exp\left[-\frac{r^2}{4(\tau+\tau_0)}\right],
\end{equation}
and $\tau_0=\sigma_{\rm inj}^2/2$. In the limit $\tau_0 \rightarrow 0$, this reduces to the case where the injection events are $\delta$ functions in space as well as time. However, this case has somewhat undesirable statistical properties (in particular, we show below that in this case the variance of the distribution diverges), and thus we will restrict our attention to small but finite $\tau_0$.

We are interested in the statistical properties of the solutions given by \autoref{eq:formal_solution} when $\mathbf{r}_i$ and $\tau_i$ are randomly distributed in space and time. To make progress in this computation, we must specify a finite domain in space and time over which injection events can occur. This is necessary because a uniform distribution over infinite space or time is not well defined. For time we use a top hat distribution: we consider injection to have begun at time 0 and continued up to the present time $\tau_f$, and we let $\tau'_i = \tau_f - \tau_i$ be the time before present for any particular injection event. Thus our time distribution is
\begin{equation}
\label{eq:pdf_time}
p_\tau(\tau') = \left\{
\begin{array}{ll}
1/\tau_f, & 0 < \tau' < \tau_f \\
0, & \mathrm{otherwise}
\end{array}
\right.
\end{equation}
In space, rather than use a similarly flat distribution, for the purposes of analytic solution it is more convenient to adopt a distribution that is not flat but that becomes so asymptotically as we consider larger domains. Specifically, we adopt a polar coordinate system and take the probability distribution of radial and angular positions to be
\begin{eqnarray}
\label{eq:pdf_r}
p_r(r) & = & \frac{2r}{R^2} e^{-(r/R)^2} \\
\label{eq:pdf_theta}
p_\theta(\theta) & = & \frac{1}{2\pi}
\end{eqnarray}
where $R$ is a free parameter and $\theta$ is in the range $[0,2\pi)$. For $r \ll R$, this is $p_r(r) = 2r/R^2$, which is the radial probability distribution function (PDF) for a distribution that is flat in space. Thus in the limit $R\rightarrow \infty$, this distribution approaches one that is flat everywhere, which is the case in which we are interested. The corresponding number of injection events per unit time per unit area is
\begin{equation}
\Gamma = \frac{2r}{R^2} e^{-(r/R)^2}
\end{equation}
so that the expected number of events over a time interval $\tau_f$ and over all space is
\begin{equation}
\lambda = \tau_f \int_0^\infty 2\pi r \Gamma \, dr = \pi R^2 \tau_f.
\end{equation}
Again, this approaches the flat distribution in which we are interested as $R\rightarrow \infty$. For this choice of domain, the expected value of the dimensionless metal density at $r=0$ is
\begin{eqnarray}
\label{eq:Sx_ex_def}
\langle S_X\rangle & = & \lambda \int_0^\infty p_w(w) \int_0^{\tau_f} p_{\tau}(\tau') 
\nonumber \\
& &
\qquad
\int_0^R p_r(r) w \phi(r,\tau')\, dr \, d\tau' \, dw
\\
& = & \frac{R^2}{4} \ln\left(1 + \frac{4\tau_f}{R^2 + 4\tau_0}\right),
\label{eq:Sx_ex}
\end{eqnarray}
where in the second step we have used the fact that, by construction, $\langle w \rangle = \int w p(w) \, dw = 1$. For large $R$, and $r \ll R$, we have
\begin{equation}
\lim_{R\rightarrow\infty} \langle S_X\rangle = \tau_f.
\end{equation}
This result is exactly as we would expect, since we have chosen to work in units where the average injection rate is one mass unit per unit area per unit time. In the case of an infinite, uniform medium the average metal abundance is simply the number of units of time for which metals have been injected.

\subsection{Variance}

We first calculate the variance of $S_X$; in terms of observables, this will set the dispersion in metallicity that we expect for stars of the same age, or the dispersion in metallicity that we expect to measure in the interstellar medium at fixed galactocentric radius. Our strategy to accomplish this is to invoke the central limit theorem. First consider a single injection event, occurring at a radius chosen from \autoref{eq:pdf_r} and a time chosen from \autoref{eq:pdf_time}. The expected value of the metal field at $r=0$ produced by this event is simply given by \autoref{eq:Sx_ex_def} evaluated with the number of events $\lambda$ set equal to unity, i.e.,
\begin{equation}
\left\langle S_X\right\rangle_1 = \frac{1}{4\pi \tau_f} \ln\left(1+\frac{4\tau_f}{R^2+4\tau_0}\right).
\end{equation}
Here we use the notation $\langle \cdot \rangle_1$ to indicate the expectation value for a single injection event. The variance for a single event is
\begin{equation}
\sigma_1^2 = \langle S_X^2\rangle_1 - \langle S_X\rangle_1^2,
\end{equation}
with
\begin{eqnarray}
\langle S_X^2\rangle_1 & = & \int_0^\infty p_w(w) \int_0^{\tau_f} p_{\tau}(\tau')
\nonumber \\
& &
\qquad
 \int_0^R p_r(r) w^2 \phi(r,\tau')^2\, dr \, d\tau' \, dw \\
& = & \frac{1+\sigma_w^2}{8\pi^2 R^2 \tau_f} \ln\left[\left(1+\frac{\tau_f}{\tau_0}\right)\frac{R^2+2\tau_0}{R^2+2(\tau_f+\tau_0)}\right],
\end{eqnarray}
where
\begin{equation}
\sigma_w^2 = \langle w^2 \rangle - \langle w \rangle^2 = \int_0^\infty w^2 p(w)\, dw - 1
\end{equation}
is the variance of $p_w(w)$. Note that the variance $\sigma_1^2$ is finite only for $\tau_0 > 0$. This justifies our earlier statement that we will only consider injection functions with $\tau_0 > 0$.

Now consider the metal density field that results from exactly $N$ injection events,
\begin{equation}
S_X = \sum_{i=1}^{N} S_{X,i},
\end{equation}
where $S_{X,i}$ is the metal surface density produced by the $i$th injection event. Since $S_X$ is the sum of independent, identically distributed variables, each drawn from a distribution with finite expectation value and variance, we can apply the central limit theorem to conclude that, as $N\rightarrow\infty$, the distribution $p_N(S_X)$ from the sum of $N$ injection events approaches a Gaussian
\begin{equation}
p_N(S_X) \approx G(\langle S_X\rangle_N, \sigma^2_N)
\end{equation}
with expectation value
\begin{equation}
\langle S_X\rangle_N = \nu \lambda \langle S_X\rangle_1 = \nu\frac{R^2}{4} \ln\left(1+\frac{4\tau_f}{R^2+4\tau_0}\right)
\end{equation}
and variance
\begin{eqnarray}
\sigma^2_N & = & \nu \lambda \sigma^2_1 
\nonumber
\\
& = & \nu \frac{1+\sigma_w^2}{8\pi}  \ln\left[\left(1+\frac{\tau_f}{\tau_0}\right)\frac{R^2+2\tau_0}{R^2+2(\tau_f+\tau_0)}\right] 
\nonumber \\
& & \qquad {}
- \nu \lambda \langle S_X\rangle_1^2.
\end{eqnarray}
where $\lambda = \pi R^2 \tau_f$ is the expected number of events, and we have defined $\nu\equiv N / \lambda$, i.e., $\nu$ is the ratio of the actual number of events to the expected number. 

Next we marginalise over the number of events $N$, which is drawn from a Poisson distribution with expectation value $\lambda$. The total probability distribution for $S_X$ is
\begin{equation}
p(S_X) = \sum_{N=1}^\infty p_N(S_X) P_\lambda(N),
\label{eq:p_SX}
\end{equation}
where $P_\lambda(N) = e^{-\lambda} \lambda^N/N!$ is the standard Poisson probability of exactly $N$ events occurring when the expectation value is $\lambda$ events. Thus $p(S_X)$ is a compound Poisson distribution. The expectation value of $\langle S_X\rangle$ for this distribution follows directly from Wald's equation, and is
\begin{equation}
\langle S_X \rangle = \frac{R^2}{4} \ln\left(1+\frac{4\tau_f}{R^2+4\tau_0}\right).
\end{equation}
Note that this expression is identical to the one we obtained by direct integration in \autoref{eq:Sx_ex}, as it should be. We can compute the variance $\sigma^2$ for $S_X$ as follows:
\begin{eqnarray}
\sigma^2 & = & \langle S_X^2\rangle - \langle S_X\rangle^2 \\
& = & \sum_{N=1}^\infty \langle S_X^2\rangle_N P_\lambda(N) - \lambda^2 \langle S_X\rangle_1^2 \\
& = & \sum_{N=1}^\infty \left(N^2 \langle S_X\rangle_1^2 + N \sigma_1^2\right) P_\lambda(N) - \lambda^2 \langle S_X\rangle_1^2 \\
& = & \lambda \sigma_1^2 + \langle S_X\rangle_1^2 \left(\sum_{N=1}^\infty N^2 P_\lambda(N) - \lambda^2\right) \\
& = & \lambda \left(\sigma_1^2 + \langle S_X \rangle_1^2\right).
\end{eqnarray}
The last step here follows from the fact that the variance of a Poisson distribution is equal to its expectation value. We therefore arrive at our final result for the variance of $S_X$,
\begin{equation}
\sigma^2 = \frac{1+\sigma_w^2}{8\pi}  \ln\left[\left(1+\frac{\tau_f}{\tau_0}\right)\frac{R^2+2\tau_0}{R^2+2(\tau_f+\tau_0)}\right].
\end{equation}
In the limit $R \rightarrow \infty$, the mean and variance approach
\begin{eqnarray}
\lim_{R\rightarrow\infty} \langle S_X\rangle & = & \tau_f \\
\lim_{R\rightarrow\infty} \sigma^2 & = & \frac{1+\sigma_w^2}{8\pi} \ln \left(1+\frac{\tau_f}{\tau_0}\right)
\end{eqnarray}
However, we caution that we have not fully characterised the distribution $p(S_X)$ because, while $p_N(S_X)$ is Gaussian in the limit $N\rightarrow \infty$, $p(S_X)$ is not. One can see this by noting that the dispersion of $p_N(S_X)$ is a function of $N$, and thus $p(S_X)$ is a sum of Gaussians with different dispersions. Such a sum is not precisely Gaussian, and thus higher moments of $p(S_X)$ are not strictly zero.

\subsection{Two Point Statistics: Correlations and Power Spectra}

\subsubsection{Spatial Correlations at Fixed Time}

We next turn to the problem of characterising the expected Pearson correlation of $S_X$, and its Fourier domain equivalent, the power spectrum. To avoid having to consider the boundaries of the injection region, we will limit ourselves to computing the correlation at scales $r \ll R$, and equivalently the power spectrum at wave numbers $k \gg 1/R$. Since we will eventually take the limit as $R\rightarrow \infty$, this is not a limitation. Consider some realisation of the metal field $S_X$, produced by drawing a particular set of injection events from the appropriate number, space, time, and mass distributions. Consider a circle of radius $R'$ and area $A = \pi R'^2$ centred on the origin. Formally, we define the correlation for this realisation evaluated for a displacement $\mathbf{r}$ on this circle as
\begin{equation}
\xi(\mathbf{r}\mid S_X) = \frac{
\overline{S_X(\mathbf{r}+\mathbf{r}') S_X(\mathbf{r}')} -
\overline{S_X}^2}{\overline{\left(S_X - \overline{S_X}\right)^2}},
\end{equation}
where for any spatial field $q$ we define
\begin{equation}
\overline{q} = \frac{1}{A}\int_A q(\mathbf{r'})\, d^2 r'
\end{equation}
as the average of $q$ over $A$. We restrict ourselves to the case $1 \ll R' \ll R$ and $|\mathbf{r}| \ll R'$. In words, we require that the averaging region be large enough to contain many injection events per unit time, but small enough that it is restricted to the region where the injection rate per unit area is approximately constant, and we only consider spatial lags that are smaller than the averaging region. We will also take the limit $R'\rightarrow \infty$, but we do so in such a way that we always have $R' \ll R$. We are interested in the expectation value of this correlation evaluated over all realisations of the metal field,
\begin{equation}
\xi(r) = \left\langle\frac{
\overline{S_X(\mathbf{r}+\mathbf{r}') S_X(\mathbf{r}')} -
\overline{S_X}^2}{\overline{\left(S_X - \overline{S_X}\right)^2}}\right\rangle,
\end{equation}
where we use angle brackets to indicate an average over realisations of $S_X$, as distinct from averages over space, as indicated by overlines. Note that, by symmetry, $\xi(r)$ must be a function of the magnitude $r = |\mathbf{r}|$ only, rather than the full vector $\mathbf{r}$. 

We can evaluate $\xi$ by noting that, in the limit of large $R$ and $R'$, we can think of the metal field $S_X$ as containing an arbitrarily large number of independent patches, each sampled from the PDF $p(S_X)$. In this limit the terms in the numerator and denominator of $\xi(r)$ are uncorrelated, so the expectation value of realisation can be applied independently to each of them. Moreover, in this case we need not distinguish between averages over realisation and averages over position, since an integral over $A$ necessarily fully samples all realisations of $S_X$, and vice versa (i.e., the metal field is ergodic). This greatly simplifies the evaluation, since it allows us to write
\begin{eqnarray}
\left\langle \overline{S_X^2}\right\rangle & \approx & \langle S_X\rangle^2 \\
\left\langle \overline{\left(S_X - \overline{S_X}\right)^2}\right\rangle & \approx & \sigma^2,
\end{eqnarray}
and the correlation reduces to
\begin{equation}
\xi(r) = \frac{\left\langle \overline{S_X(\mathbf{r}+\mathbf{r}') S_X(\mathbf{r}')}\right\rangle
- \langle S_X\rangle^2}{\sigma^2}.
\label{eq:pearson_corr}
\end{equation}

We can approximate the remaining term by
\begin{eqnarray}
\lefteqn{
\left\langle \overline{S_X(\mathbf{r}+\mathbf{r}') S_X(\mathbf{r}')}\right\rangle =
} \nonumber \\
& &
\frac{1}{A} \int \left\langle S_{X,A}(\mathbf{r}+\mathbf{r}') S_{X,A}(\mathbf{r'})\right\rangle \, d^2 r',
\label{eq:corr_int_approx}
\end{eqnarray}
where
\begin{equation}
\label{eq:SXA}
S_{X,A}(\mathbf{r'}) \equiv \sum_{i, r_i<R', \tau_i>0} w_i \phi(\mathbf{r'}-\mathbf{r}_i,\tau_i),
\end{equation}
and for convenience we define $\tau_i$ as the present time minus the time at which injection $i$ occurred. This approximation is non-trivial, and does not simply follow from ergodicity, because we have limited the sum to include only those events inside $A$, and we have changed the area of integration to be over all space rather than simply over $A$. In words, our approximation is that the correlation of the entire metal field $S_X$ averaged over the region $A$ can be approximated by the correlation of the field $S_{X,A}$ that would be produced solely by injection events that lie within $A$ (and that occurred before the present), integrated over all space rather than simply over $A$. The replacement of an integral over $A$ by an integral over all space is easily justified, since we are taking the limit $R'\rightarrow \infty$. It is somewhat less immediately obvious that we can only consider events inside $A$, and neglect those outside it. We demonstrate that this is in fact the case in \autoref{app:proof}.
 
Having justified our approximation, the problem is now much simpler, because we can evaluate \autoref{eq:corr_int_approx} with the aid of the Wiener-Khinchin Theorem. This requires that
\begin{eqnarray}
\lefteqn{
\frac{1}{A} \int \left\langle S_{X,A}(\mathbf{r}+\mathbf{r'}) S_{X,A}(\mathbf{r}')\right\rangle\, d^2 r'
=  \int \Psi(k) e^{-i\mathbf{k}\cdot\mathbf{r}}\, d^2 k 
}
\label{eq:wiener-khinchin1}
\\
 & = & 2 \pi \int_0^\infty \Psi(k) J_0(k r) k \, dk
 \label{eq:wiener-khinchin}
\end{eqnarray}
where the integrals run over all real and Fourier space, $J_n(x)$ is the Bessel function of the first kind of order $n$,
\begin{equation}
\Psi(k) \equiv \frac{1}{A} \left\langle\left|\tilde{S}_{X,A}(\mathbf{k})\right|^2\right\rangle
\end{equation}
is the expected area-normalised power spectral density, and
\begin{equation}
\tilde{S}_{X,A}(\mathbf{k}) = \frac{1}{2\pi} \int S_{X,A}(\mathbf{r'})e^{i\mathbf{k}\cdot\mathbf{r'}}\, d^2 r'
\end{equation}
is the Fourier transform of $S_{X,A}$. Note that by symmetry $\Psi(k)$ must be a function of $k = |\mathbf{k}|$ only, and \autoref{eq:wiener-khinchin} follows from this lack of angular dependence.

The Fourier transform is
\begin{equation}
\tilde{S}_{X,A}(\mathbf{k}) = \sum_{i,r_i<R', \tau_i>0} w_i \tilde{\phi}(\mathbf{k}, \mathbf{r}_i, \tau_i)
\end{equation}
where 
\begin{equation}
\tilde{\phi}(\mathbf{k}, \mathbf{r}_i, \tau_i) = \frac{1}{2\pi} \exp\left[-\left(\tau_i+\tau_0\right)k^2 + i \mathbf{k}\cdot\mathbf{r}_i\right].
\end{equation}
Since $w$, $\mathbf{r}$, and $\tau$ are all independent random variables, the expected, normalised power spectral density is therefore
\begin{eqnarray}
\lefteqn{\Psi(k)=\frac{e^{-2\tau_0 k^2}}{4\pi^3 R'^2} \sum_i 
\left\{
\left\langle w_i^2\right\rangle
\left\langle e^{-2\tau_i k^2}\right\rangle + {}
\phantom{
\sum_{j\neq i} 
\left\langle w_i w_j\right\rangle
\left\langle e^{-(\tau_i+\tau_j)k^2}\right\rangle \left\langle \cos\left[\mathbf{k}\cdot(\mathbf{r}_i-\mathbf{r}_j)\right]\right\rangle
\hspace{-3in}}
\right.
}
\nonumber \\
& & 
\left.
\sum_{j\neq i} 
\left\langle w_i w_j\right\rangle
\left\langle e^{-(\tau_i+\tau_j)k^2}\right\rangle \left\langle \cos\left[\mathbf{k}\cdot(\mathbf{r}_i-\mathbf{r}_j)\right]\right\rangle\right\},
\label{eq:sx2}
\end{eqnarray}
where the first sum includes all events at times $\tau_i > 0$ and radii $r_i < R'$, and the second includes all events satisfying those conditions and with index $j \neq i$; for the sake of brevity we shall from this point forward omit the conditions on $i$ and $j$. In \autoref{eq:sx2}, the first term inside curly brackets represents the correlation of a single event with itself, while the second term represents the cross-correlation between different injection events. 

We next compute the expected power spectral density, marginalised over the PDF of injection event numbers, masses, times, and locations. First consider a fixed number of injection events $N$. The expected power spectral density is
\begin{eqnarray}
\lefteqn{\Psi(k,N) = \frac{e^{-2\tau_0 k^2}}{4\pi^3 R'^2} N
\left[
\left\langle w_i^2\right\rangle
\left\langle e^{-2\tau_i k^2}\right\rangle + {}
\right.
}
\nonumber \\
& &
\left.
(N-1)
\left\langle w_i w_j\right\rangle
\left\langle e^{-(\tau_i+\tau_j) k^2}\right \rangle
\left\langle \cos\left[\mathbf{k}\cdot(\mathbf{r_i}-\mathbf{r_j})\right]\right\rangle\right].
\end{eqnarray}
It is straightforward to evaluate each of the individual expectation values in the expression for $\Psi(k,N)$. The averages over $w$ are simply $\langle w_i^2\rangle = 1 + \sigma_w^2$ and $\langle w_i w_j \rangle = \langle w_i \rangle \langle w_j\rangle = 1$. The next term is
\begin{equation}
\left\langle e^{-2\tau_i k^2}\right\rangle = \int_0^\infty e^{-2\tau k^2} p_\tau(\tau) \, d\tau
= \frac{1 - e^{-2 \tau_f k^2}}{2 \tau_f k^2}
\label{eq:expec_t}
\end{equation}
For the term involving $\tau_i + \tau_j$, note that the PDF for the sum of two times $\tau_2 = \tau_i + \tau_j$ that are each drawn from $p_\tau(\tau)$ is given by the autoconvolution of $p_\tau(\tau)$, which is
\begin{equation}
p_{\tau_2}(\tau_2) = 
\frac{1}{\tau_f^2}
\left\{
\begin{array}{ll}
\tau_2, & 0 \leq \tau_2 < \tau_f \\
2 \tau_f - \tau_2, & \tau_f \leq \tau_2 < 2\tau_f \\
0, & \mathrm{otherwise}
\end{array}
\right..
\end{equation}
Thus
\begin{equation}
\left\langle e^{-(\tau_i+\tau_j) k^2}\right \rangle =
\int_0^{\infty} e^{-\tau_2 k^2} p_{\tau_2}(\tau_2) \, d\tau_2
= \left(\frac{1-e^{-\tau_f k^2}}{\tau_f k^2}\right)^2.
\label{eq:expec_t2}
\end{equation}
Finally, for the expectation value $\langle \cos\left[\mathbf{k}\cdot(\mathbf{r}_i-\mathbf{r}_j)\right]\rangle$, it is helpful to define $\Delta r = |\mathbf{r}_i-\mathbf{r}_j|$ and $\theta = \mbox{arg}(\mathbf{r}_i - \mathbf{r}_j-\mathbf{k})$, so that
\begin{eqnarray}
\lefteqn{\left\langle \cos\left[\mathbf{k}\cdot(\mathbf{r}_i - \mathbf{r}_j)\right]\right\rangle
=
}
\nonumber \\
& &
 \int_0^\infty \int_0^{2\pi} \cos\left( k \Delta r \sin \theta\right) p_\theta(\theta) \, d\theta\, p_{\Delta r}(\Delta r) \, d\Delta r.
\end{eqnarray}
Clearly by symmetry the PDF of angles $\theta$ must be uniformly distributed, so that $p_\theta(\theta) = 1/2\pi$ and performing the integration yields
\begin{equation}
\label{eq:exp_rdiff}
\left\langle \cos\left[\mathbf{k}\cdot(\mathbf{r}_i - \mathbf{r}_j)\right]\right \rangle
= \int_0^\infty J_0(k \Delta r) p_{\Delta r}(\Delta r)\,d\Delta r.
\end{equation}
Finally, the distribution of separations between two randomly-selected points in a disc of radius $R'$ is \citep{garcia-pelayo05a}
\begin{equation}
p_{\Delta r}(\Delta r) =
\frac{4\Delta r}{\pi R'^2}
\left\{
\begin{array}{ll}
\arccos\frac{\Delta r}{2R'} -
\frac{\Delta r}{2R'} \sqrt{1-\frac{\Delta r^2}{4R'^2}}, & r < 2 R' \\
0, & \mbox{otherwise}
\end{array}
\right.
\end{equation}
where the $\arccos$ function is chosen to have a range $[0,\pi]$. Using this value of $p_{\Delta r}$ in \autoref{eq:exp_rdiff} gives
\begin{equation}
\label{eq:expec_r}
\left\langle \cos\left[\mathbf{k}\cdot(\mathbf{r}_i - \mathbf{r}_j)\right]\right \rangle = 
\left[2 \frac{J_1(kR')}{k R'}\right]^2.
\end{equation}

Inserting the expectation values of the various terms into \autoref{eq:sx2}, we obtain the expected power spectral density $\Psi(k,N)$ for a fixed number of events $N$:
\begin{eqnarray}
\lefteqn{
\Psi(k, N) = \frac{N(1+\sigma_w^2)}{8\pi^3 R'^2} e^{-2\tau_0 k^2}
\left(\frac{1-e^{-\tau_f k^2}}{\tau_f k^2}\right)
\left\{1 + {}
\phantom{\left[\frac{1 - \left(1+\tau_f k^2\right) e^{-\tau_f k^2}}{\tau_f k^2}\right]\hspace{-3in}}
\right.
}
\nonumber \\
& &
\left.
e^{-\tau_f k^2}  + 8 \frac{N-1}{1+\sigma_w^2}
 \left(\frac{1-e^{-\tau_f k^2}}{\tau_f k^2}\right)
\left[\frac{J_1(k R')}{k R'}\right]^2
\right\}.
\end{eqnarray}

The final step is to marginalise over $N$. Recall that $N$ is Poisson-distributed with expectation value $\langle N \rangle = \pi R'^2 \tau_f$. The expectation value of $N^2$ is $\langle N^2 \rangle = \langle N\rangle ^2 + \sigma_N^2$, where $\sigma_N^2$ is the variance in $N$; for a Poisson distribution the variance is equal to the expectation value, so $\langle N^2\rangle = \langle N \rangle^2 + \langle N \rangle$. Using these results, we find that the power spectral density is
\begin{eqnarray}
\lefteqn{
\Psi(k) = \frac{1+\sigma_w^2}{8\pi^2} e^{-2\tau_0 k^2}
\left(\frac{1-e^{-\tau_f k^2}}{k^2}\right)
\left\{1 + e^{-\tau_f k^2} 
+ {}
\phantom{\left[\frac{1 - \left(1+\tau_f k^2\right) e^{-\tau_f k^2}}{\tau_f k^2}\right]\hspace{-3in}}
\right.
}
\nonumber \\
& &
\qquad
\left.
\frac{8\pi}{1+\sigma_w^2} \left(\frac{1 - e^{-\tau_f k^2}}{ k^2}\right)
\left[\frac{J_1(k R')}{k}\right]^2
\right\}.
\label{eq:psd_full}
\end{eqnarray}
In the limit $R' \rightarrow \infty$ at fixed $k$, the Bessel function $J_1(k R')$ has an envelope whose magnitude scales as $R'^{-1/2}$, and so this term becomes negligible in comparison to the other terms inside the curly braces. The area-normalised power spectral density therefore approaches
\begin{equation}
\Psi(k) = \frac{1+\sigma_w^2}{8\pi^2 k^2} e^{-2\tau_0 k^2} \left(1-e^{-2 \tau_f k^2}\right).
\label{eq:psd}
\end{equation}
This expression is valid in the limit $k R' \gg 1$, and thus is valid at any finite $k$ in the limit $R' \rightarrow \infty$.

Substituting \autoref{eq:psd_full} into the Wiener-Khinchin Theorem (\autoref{eq:wiener-khinchin}), and taking the limit $R'\rightarrow \infty$, the correlation is\footnote{One important subtlety in evaluating the correlation is that we cannot use \autoref{eq:psd} for the area-normalised power spectral density because this expression is only valid in the limit $k R' \gg 1$, and the integral goes all the way to $k=0$. We must therefore use the full expression given by \autoref{eq:psd_full}.}
\begin{eqnarray}
\lefteqn{\xi(r) = \frac{2}{\left(1+\sigma_w^2\right)\ln\left(1+\tau_f/\tau_0\right)}
}
\nonumber \\
& &
\left[\left(1+\sigma_w^2\right)
\int_0^\infty e^{-2\tau_0 k^2} \left(1-e^{-2 \tau_f k^2}\right) \frac{J_0(k r)}{k}\, dk + {}
\right.
\nonumber \\
& &
\left.
\lim_{R'\rightarrow\infty} 8\pi \int_0^\infty \frac{e^{-2\tau_0 k^2}}{k^5}\left(1-e^{-\tau_f k^2}\right)^2 J_0(k r) J_1(k R')^2\, dk
\right.
\nonumber \\
& &
\left.
{} - 4\pi \tau_f^2
\phantom{\int_0^\infty \frac{e^{-2\tau_0 k^2}}{k^5}\hspace{-1.7cm}}
\right]
\end{eqnarray}
To evaluate the middle term in the square brackets, note that as $R'\rightarrow \infty$, the factor $J_1(k R')$ goes to zero except in an infinitesimally small region near $k=0$, where it produces a $\delta$ function-like spike. We can therefore evaluate the integral by expanding the prefactor preceding $J_1(k R')$ in a series about $k=0$. This gives
\begin{eqnarray}
\lefteqn{\xi(r) = \frac{2}{\left(1+\sigma_w^2\right)\ln\left(1+\tau_f/\tau_0\right)}
}
\nonumber \\
& &
\left[
\left(1+\sigma_w^2\right)
\int_0^\infty e^{-2\tau_0 k^2} \left(1-e^{-2 \tau_f k^2}\right) \frac{J_0(k r)}{k}\, dk + {}
\right.
\nonumber \\
& &
\left.
\lim_{R'\rightarrow\infty} 8\pi \tau_f^2 \int_0^\infty \frac{J_1(k R')^2}{k}\, dk
-4\pi \tau_f^2
\right]
\end{eqnarray}
\green{Finally, $\int_0^\infty J_1(kR')^2/k \, dk = 1/2$ regardless of the value of $R'$}, so the final two terms in the square brackets cancel. This gives our final expression for the correlation,
\begin{equation}
\xi(r) = \frac{2}{\ln\left(1+\tau_f/\tau_0\right)} 
\int_0^\infty e^{-2\tau_0 k^2} \left(1-e^{-2 \tau_f k^2}\right) \frac{J_0(k r)}{k}\, dk.
\label{eq:xi_final}
\end{equation}
Note that $\xi(r)$ is independent of $\sigma_w$. Thus the dimensionless correlation function does not depend on the distribution of injection event masses.

\begin{figure}
\includegraphics[width=\columnwidth]{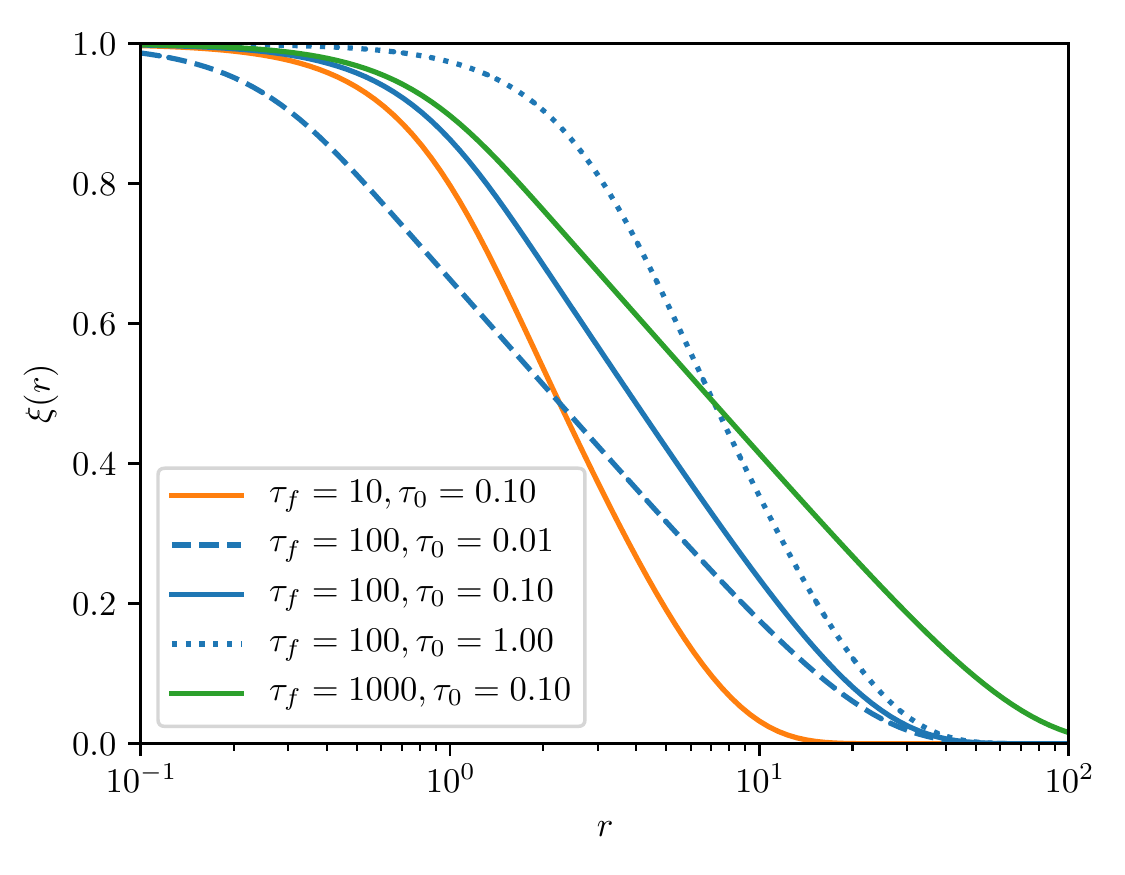}
\caption{
\label{fig:xi_const_time}
Pearson correlation of the metal field $\xi(r)$ as a function of dimensionless length $r$, evaluated for a range of star formation times $\tau_f$ and injection widths $\tau_0$.
}
\end{figure}

\begin{figure}
\includegraphics[width=\columnwidth]{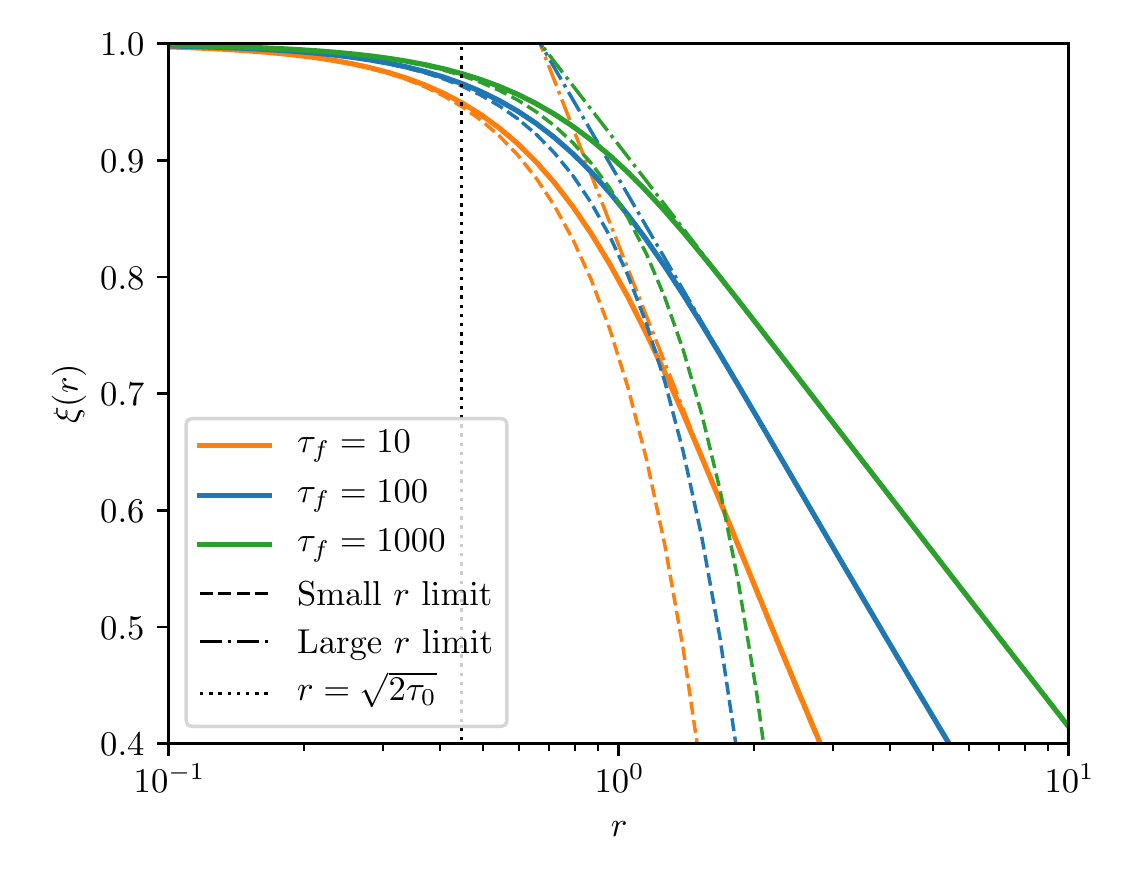}
\caption{
\label{fig:xi_analytic_approx}
Comparison of the exact, numerically-calculated Pearson correlation of the metal field $\xi(r)$ (solid lines, \autoref{eq:xi_final}) and the analytic approximations appropriate to the small $r$ (dashed lines, \autoref{eq:xi_small_r}) and large $r$ (dot-dashed lines, \autoref{eq:xi_large_r}) limits. Line colours denote the value of $\tau_f$ indicated in the legend, and all the lines shown use $\tau_0 = 0.1$. Note that the solid lines are identical to the corresponding ones shown in \autoref{fig:xi_const_time}, but the plot range has been reduced to zoom in on the region of maximum difference between the exact and approximate analytic results. The vertical dotted black line shows $r = \sqrt{2\tau_0}$, the approximate value that divides the small and large $r$ limits.
}
\end{figure}

The integral in \autoref{eq:xi_final} cannot be evaluated in closed form for arbitrary $r$ and $\tau_0$, but it is straightforward to evaluate numerically. We plot the correlation function for an astrophysically-relevant range of $\tau_f$ and $\tau_0$ values (see \autoref{ssec:phys_scales}) in \autoref{fig:xi_const_time}. We can also give a closed-form expression for correlation function at both small and large $r$. Note that the factor in front of the Bessel function in \autoref{eq:xi_final} imposes an exponential cutoff at $k \sim 1/\sqrt{2\tau_0}$, so the value of the integral is essentially determined by the behaviour at $k$ below this value. If $r \ll \sqrt{2\tau_0}$, then the integrand is only significant in locations where the argument of the Bessel function is near 0. We can therefore expand the Bessel function about 0, which yields an integrable expression. The result is
\begin{equation}
\xi(r) = 1 - \frac{r^2}{8 \tau_0(1+\tau_0/\tau_f) \ln(1 + \tau_f/\tau_0)}
\qquad(r \ll \sqrt{2\tau_0}).
\label{eq:xi_small_r}
\end{equation}
Note that $\xi(r) \rightarrow 1$ as $r\rightarrow 0$, as expected. Conversely, the Bessel function goes to zero for large arguments, so if $r$ is very large then the integrand is non-zero only for very small $k$. We can then expand $e^{-\tau_0 k^2}$ about $k=0$, again producing an integrable expression and yielding
\begin{equation}
\xi(r) = \frac{\Gamma\left(0,r^2/8\tau_f\right)}{\ln(1+\tau_f/\tau_0)}
\qquad(r \gg \sqrt{2\tau_0}),
\label{eq:xi_large_r}
\end{equation}
where $\Gamma(a,z)$ is the upper incomplete $\Gamma$ function.\footnote{\green{$\Gamma(a,z)$ here should not be confused with the event rate $\Gamma$ introduced above. To minimise confusion, we always write out the upper incomplete $\Gamma$ function with its arguments, as $\Gamma(a,z)$, while the event rate has no arguments.}} We compare the numerically-calculated function to the two limiting cases in \autoref{fig:xi_analytic_approx}.

\subsubsection{Correlations in Space and Time}

The calculation presented in the previous section can be generalised to compute the correlation of the metal in time as well as space. We consider the same spatial setup as in the previous section, but now consider metal fields measured at two different times separated by $\Delta \tau$. We will compute the correlation between the metal field as it is now, at time $\tau_f$, and as it was at an earlier time $\tau_f-\Delta \tau$. The Pearson correlation, again assuming ergodicity of the metal field, is
\begin{eqnarray}
\lefteqn{
\xi(r,\Delta \tau) = 
}
\nonumber \\
& &
\hspace{-0.5cm}
\frac{\left\langle\overline{S_X(\mathbf{r}+\mathbf{r}', \tau-\Delta \tau)
S_X(\mathbf{r}', \tau)}\right\rangle 
- \left\langle S_X\right\rangle_{\tau_f-\Delta\tau}\left\langle S_X\right\rangle_{\tau_f}}
{\sigma_{\tau_f-\Delta\tau}\sigma_{\tau_f}},
\label{eq:pearson_time}
\end{eqnarray}
where we have defined
\begin{eqnarray}
\left\langle S_X\right\rangle_{\tau} & = & \left\langle S_X(\mathbf{r}',\tau)\right\rangle
\\
\sigma_{\tau}^2 & = & \left\langle \left[S_X(\mathbf{r}',\tau)-\left\langle S_X(\mathbf{r}',\tau)\right\rangle\right]^2\right\rangle
\end{eqnarray}
as the average and variance of the metal field at time $\tau_f$, and similarly for time $\tau_f-\Delta\tau$. Note that, as expected, this definition reduces to \autoref{eq:pearson_corr} for $\Delta\tau = 0$. Using the same approach as in the previous section to evaluate the correlation integral in \autoref{eq:pearson_time}, we have
\begin{eqnarray}
\lefteqn{
\frac{1}{A} \int \left\langle S_{X,A}(\mathbf{r}+\mathbf{r}',\tau-\Delta \tau) S_{X,A}(\mathbf{r}',\tau)\right\rangle d^2 r' }
\nonumber \\
& = & 2\pi \int_0^\infty \Psi(k,\Delta \tau) J_0(k r) k \, dk
\end{eqnarray}
where
\begin{equation}
\Psi(k,\Delta\tau) = \frac{1}{A}\left\langle \tilde{S}_{X,A}(\mathbf{k},\tau-\Delta\tau)  \tilde{S}_{X,A}^*(\mathbf{k},\tau) \right\rangle
\end{equation}
is the area-normalised power spectral density for the cross correlation; here the asterisk denotes complex conjugation. The power spectral density in turn is
\begin{eqnarray}
\lefteqn{\Psi(k,\Delta \tau)=\frac{e^{(\Delta \tau-2\tau_0) k^2}}{4\pi^3 R'^2} \sum_{i}^{\tau_i>\Delta \tau} 
\left\{
\left\langle w_i^2\right\rangle
\left\langle e^{-2\tau_i k^2}\right\rangle
+ {}
\phantom{2 \sum_{j>i}^{\tau_j>0, \tau_i > \Delta \tau} \left\langle e^{-(\tau_i+\tau_j)k^2}\right\rangle \hspace{-3in}}
\right.
}
\nonumber \\
& & 
\hspace{-0.2in}
\left.
\sum_{j\neq i}^{\tau_j>0, \tau_i > \Delta \tau}
\left\langle w_i w_j \right\rangle
\left\langle e^{-(\tau_i+\tau_j)k^2}\right\rangle \left\langle \cos\left[\mathbf{k}\cdot(\mathbf{r}_i-\mathbf{r}_j)\right]\right\rangle\right\},
\label{eq:sx2_time}
\end{eqnarray}
where we define $\tau_i$ as the time elapsed between injection $i$ and the present. For compactness we have omitted the condition $r_i < R'$ on the sums, but they should be understood to include only those injection events that are located with $A$.

Note that \autoref{eq:sx2_time} is nearly identical to \autoref{eq:sx2}. The only difference is the presence of an extra factor $e^{\Delta \tau k^2}$. The first term in curly brackets, representing correlations of events with themselves, only includes events that occurred at least a time $\Delta \tau$ in the past and thus contribute to the metal field both now and at the time $\Delta\tau$ in the past that we are considering; the second term in curly brackets, representing the cross-correlation between events, includes both events that occurred longer than $\Delta \tau$ before the present, and the other events that occurred at any time before present.

Let $N$ be the number of injection events up to the present time $\tau$, and let $\Delta N$ be the number of events that occurred between time $\tau-\Delta\tau$ and time $\tau$. The normalised power spectral density for these numbers of events is
\begin{eqnarray}
\lefteqn{\Psi(k,N,\Delta N) = \frac{e^{(\Delta\tau-2\tau_0) k^2}}{4\pi^3 R'^2} (N-\Delta N)
\left[
\left\langle w_i^2\right\rangle
\left\langle e^{-2\tau_i k^2}\right\rangle 
+ {}
\right.
}
\nonumber \\
& &
\left.
(N-1)
\left\langle w_i w_j\right\rangle
\left\langle e^{-(\tau_i+\tau_j) k^2}\right \rangle
\left\langle \cos\left[\mathbf{k}\cdot(\mathbf{r_i}-\mathbf{r_j})\right]\right\rangle
\right].
\label{eq:psi_k_n_dn}
\end{eqnarray}
Evaluation of the individual correlation terms proceeds much as in the fixed-time case. The expectation values over $w$ and $\mathbf{r}$ are unchanged. The distribution of times $i$ in the first angle bracket term is
\begin{equation}
p_\tau(\tau_i) =
\left\{
\begin{array}{ll}
1/(\tau_f-\Delta\tau), & \Delta\tau < \tau_i < \tau_f \\
0, & \mathrm{otherwise}
\end{array}
\right.,
\label{eq:pdf_time_mod}
\end{equation}
i.e., the events are uniformly distributed in time between times $\Delta\tau$ and $\tau_f$ in the past. Thus we have
\begin{equation}
\left\langle e^{-2\tau_i k^2}\right\rangle = \frac{e^{-2 \Delta\tau k^2} - e^{-2\tau_f k^2}}{2 (\tau_f-\Delta \tau)k^2}.
\end{equation}
Similarly, the sum of the two times $\tau_i + \tau_j$ is the sum of one number drawn from the distribution given by \autoref{eq:pdf_time_mod} with another drawn from a uniform distribution between 0 and $\tau_f$ (\autoref{eq:pdf_time}). The PDF for the sum is the convolution of the PDFs for $\tau_i$ and $\tau_j$, which is
\begin{equation}
p_{\tau_2}(\tau_2) =
\frac{1}{\tau_f(\tau_f-\Delta\tau)}
\left\{
\begin{array}{ll}
\tau_2-\Delta\tau, & \Delta\tau < \tau_2 \leq \tau_f \\
\tau_f -\Delta\tau, & \tau_f < \tau_2 \leq \tau_f + \Delta \tau \\
2\tau_f - \tau_2, & \tau_f+\Delta\tau < \tau_2 \leq 2\tau_f \\
0, & \mathrm{otherwise}
\end{array}
\right.
\end{equation}
Thus the second angle bracket over time is
\begin{equation}
\left\langle e^{-(\tau_i+\tau_j)k^2}\right\rangle
= \frac{\left(1-e^{-\tau_f k^2}\right)\left(e^{-\Delta\tau k^2}-e^{-\tau_f k^2}\right)}{\tau_f\left(\tau_f-\Delta\tau\right) k^4}.
\end{equation}
Inserting the expectation values into \autoref{eq:psi_k_n_dn} gives
\begin{eqnarray}
\lefteqn{\Psi(k,N,\Delta N) = \left(\frac{N-\Delta N}{8\pi^3 R'^2}\right) e^{-2\tau_0 k^2} 
\left[ \frac{1 - e^{(\Delta\tau-\tau_f) k^2}}{(\tau_f-\Delta \tau) k^2}\right]
}
\nonumber \\
& &
\left\{
\left(1+\sigma_w^2\right)
\left(e^{-\Delta \tau k^2} + e^{-\tau_f k^2} \right)
\phantom{ \left[\frac{J_1(k R')}{k R'}\right]^2 }
\right.
\nonumber \\
& &
\left.
\quad
{} + 
8 (N-1) \left(\frac{1 - e^{-\tau_f k^2}}{\tau_f k^2}\right) \left[\frac{J_1(k R')}{k R'}\right]^2\right\}
\label{eq:psi_k_n_dn_eval}
\end{eqnarray}

Finally, we must marginalise over the numbers of events. This requires that we compute the expectation values $\langle N\rangle$, $\langle\Delta N\rangle$, $\langle N^2\rangle$, and $\langle N\Delta N\rangle$. As in the previous case, $N$ is Poisson-distributed with expectation value $\langle N \rangle = \pi R'^2 \tau_f$, and thus $\langle N^2\rangle = \langle N\rangle^2 + \langle N \rangle$. Similarly, $\Delta N$ is Poisson-distributed with expectation value $\langle \Delta N\rangle = \pi R'^2 \Delta \tau$. The expectation value $\langle N\Delta N\rangle$ is slightly subtle because $N$ and $\Delta N$ are not independent quantities, since the number of events that occurred in the most recent interval of $\Delta \tau$ contributes to the total number that occurred over time $\tau_f$. However, we can evaluate this term by letting $N_1 = N - \Delta N$ represent the number of events that occurred between $\tau_f$ and $\tau_f - \Delta\tau$ ago, and writing $N \Delta N = N_1 \Delta N + \Delta N^2$. The advantage of this expression is that, since the time intervals $\tau_f$ to $\tau_f - \Delta \tau$ and $\Delta \tau$ to 0 are disjoint and the injection events are uncorrelated in time, $N_1$ and $\Delta N$ are independent, and thus the expectation value of their product is just the product of their expectation values. We therefore have
\begin{eqnarray}
\langle N\Delta N \rangle & = & \left\langle N_1\right\rangle\left\langle \Delta N\right\rangle + \left\langle \Delta N^2\right\rangle \\
& = & \left(\left\langle N\right\rangle - \left\langle\Delta N\right\rangle\right) \left\langle \Delta N \right\rangle + \left\langle \Delta N^2\right\rangle \\
& = & \left\langle N\right\rangle \left\langle\Delta N \right\rangle + \left\langle \Delta N^2\right\rangle - \left\langle \Delta N\right\rangle^2 \\
& = & \left\langle \Delta N\right\rangle \left( \left\langle N\right\rangle + 1\right),
\end{eqnarray}
where the last step follows from the fact that $\Delta N$ is Poisson-distributed, and thus has a variance $\left\langle \Delta N^2\right\rangle - \left\langle \Delta N\right\rangle^2$ equal to its expectation value. Using the various number expectation values in \autoref{eq:psi_k_n_dn_eval} gives
\begin{eqnarray}
\lefteqn{
\Psi(k,\Delta\tau) = \frac{1+\sigma_w^2}{8\pi^2} e^{-2\tau_0 k^2}
\left[ \frac{1 - e^{(\Delta\tau-\tau_f) k^2}}{k^2}\right]
}
\nonumber \\
& &
\left\{
e^{-\Delta \tau k^2} + e^{-\tau_f k^2}
+ {}
\phantom{ \left[\frac{J_1(k R')}{k R'}\right]^2 \hspace{-4in}}
\right.
\nonumber \\
& & 
\quad
\frac{8 \pi}{1+\sigma_w^2} \left[1 + \frac{2\Delta \tau}{\pi R'^2 \tau_f(\tau_f-\Delta \tau)}\right] \cdot {}
\nonumber \\
& &
\left.
\quad
\left(\frac{1 - e^{-\tau_f k^2}}{k^2}\right)
\left[\frac{J_1(k R')}{k}\right]^2
\right\}.
\end{eqnarray}
In the limit $R' \rightarrow \infty$, for any $k>0$, this reduces to
\begin{equation}
\Psi(k,\Delta\tau) =
\frac{1+\sigma_w^2}{8\pi^2 k^2}
e^{-(\Delta\tau+2\tau_0)k^2}
\left[1-e^{-2(\tau_f-\Delta\tau)k^2}\right].
\label{eq:psd_time}
\end{equation}
This expression has the limiting behaviour we expect: for $\Delta\tau = \tau_f$ the power goes to zero, since in the initial time in this case has no metals present, while for $\Delta\tau = 0$ \autoref{eq:psd_time} reduces to \autoref{eq:psd}, the power spectrum for the metal field at fixed time.

\begin{figure}
\includegraphics[width=\columnwidth]{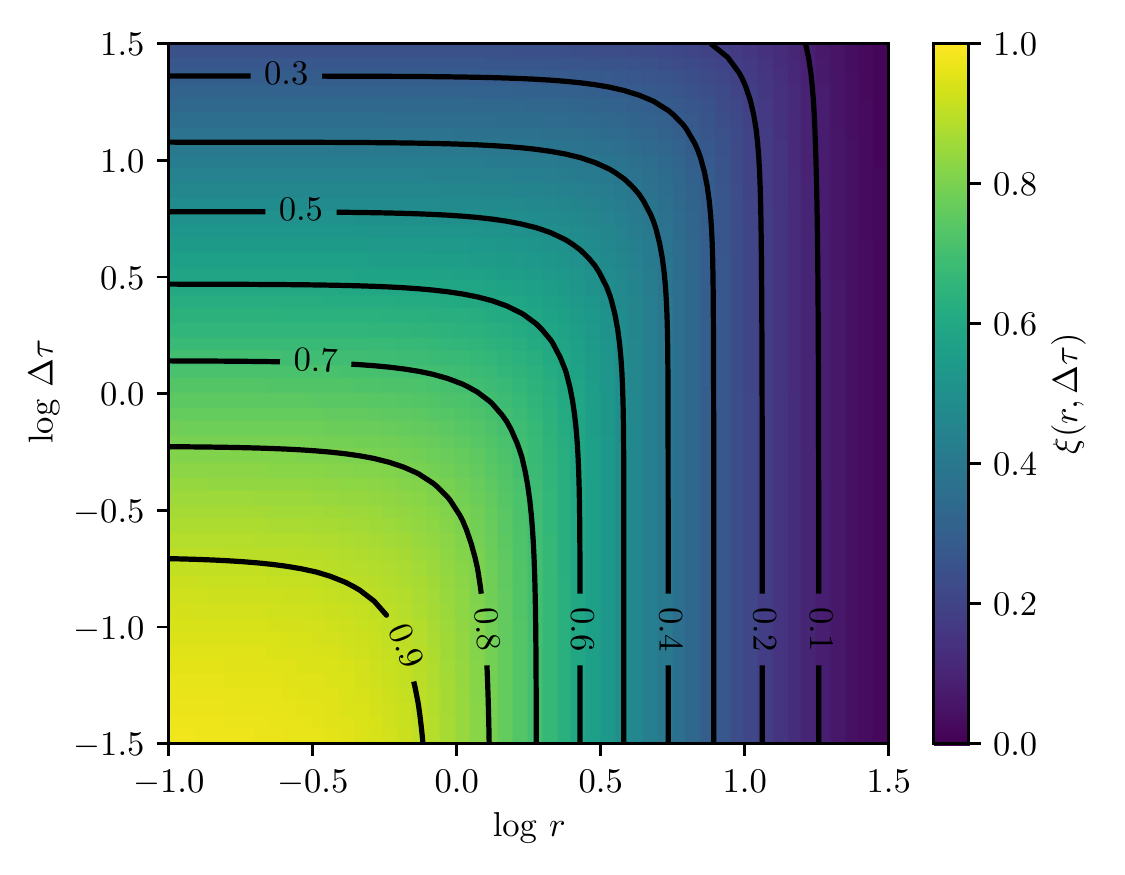}
\caption{
\label{fig:xi_var_time}
Perason correlation of the metal field $\xi(r,\Delta\tau)$ as a function of dimensionless length $r$ and time difference $\Delta \tau$. The figure shown is for dimensionless star formation time $\tau_f = 100$ and injection with $\tau_0 = 0.1$.
}
\end{figure}

Again taking the limit $R'\rightarrow \infty$, and evaluating the $J_1(k R')$ integral by series expansion about $k=0$ as in the previous section, we obtain the correlation
\begin{eqnarray}
\lefteqn{
\xi(r,\Delta\tau) = 
\frac{2}{\sqrt{\ln\left(1+\frac{\tau_f}{\tau_0}\right)\ln\left(1+\frac{\tau_f-\Delta\tau}{\tau_0}\right)}}
\cdot {}
}
\nonumber \\
& &
\int_0^\infty e^{-(\Delta\tau+2\tau_0)k^2}
\left[1-e^{-2(\tau_f-\Delta\tau)k^2}\right] \frac{J_0(k r)}{k}\, dk.
\label{eq:xi_t_final}
\end{eqnarray}
As with the correlation at a single time, the integral cannot be evaluated in closed form for arbitrary arguments, but is straightforward to evaluate numerically, and to evaluate analytically in limiting cases. For $r \ll \sqrt{2\tau_0+\Delta\tau}$, expanding $J_0(k r)$ about $r=0$, we have
\begin{equation}
\xi(r,\Delta\tau) \approx \frac{
\ln\left[\frac{2(\tau_f+\tau_0)-\Delta\tau}{2\tau_0+\Delta\tau}\right]
- r^2 \frac{1-\Delta\tau/\tau_f}{2(2\tau_0+\Delta\tau)[2(1+\tau_0/\tau_f)-\Delta\tau/\tau_f]}
}
{\sqrt{\ln\left(1+\frac{\tau_f}{\tau_0}\right)\ln\left(1+\frac{\tau_f-\Delta\tau}{\tau_0}\right)}}.
\end{equation}
Similarly, for $r \gg \sqrt{2\tau_0+\Delta\tau}$ we can expand the first exponential factor in the integral about $k=0$ to obtain
\begin{equation}
\xi(r,\Delta\tau) \approx
\frac{\Gamma(0,r^2/8(\tau_f-\Delta\tau))}{\sqrt{\ln\left(1+\frac{\tau_f}{\tau_0}\right)\ln\left(1+\frac{\tau_f-\Delta\tau}{\tau_0}\right)}}.
\end{equation}
We plot a sample numerical evaluation of $\xi(r,\Delta\tau)$ in \autoref{fig:xi_var_time}.

\section{Astrophysical Implications}
\label{sec:implications}

Having derived the statistical properties of our formal system, we are now in a position to explore the astrophysical implications of our findings.

\subsection{Parameter Estimates}
\label{ssec:phys_scales}

As a first step to working out the astrophysical implications of our model, we must estimate characteristic astrophysical values for various dimensional and dimensionless parameters that enter our solution. For convenience we collect typical parameter values for the Milky Way near the Solar Circle in \autoref{tab:mw_params}.

\begin{table*}
\begin{tabular}{ccc@{\qquad}cccc}
\hline
& & & \multicolumn{4}{c}{Origin Sites} \\
Parameter & Units & Description & SNII & SNIa & AGB & NSM \\ \hline
\multicolumn{3}{c}{ISM and star formation parameters} \\ \hline
$t_*$ & Gyr & Star formation duration & \multicolumn{4}{c}{10} \\
$\kappa$ & pc km s$^{-1}$ & Diffusion coefficient & \multicolumn{4}{c}{300}  \\
$\Gamma$ & pc$^{-2}$ Myr$^{-1}$ & Event rate & $3 \times 10^{-6}$ & $3\times 10^{-4}$ & $5\times 10^{-4}$ & $10^{-7}$ \\ 
$\sigma_w^2$ & - & Injection mass variance & 20 & 0.1 & 5 & 10? \\ \hline
\multicolumn{3}{c}{Length and time scales} \\ \hline
$x_s$ & pc & Length scale & 100 & 32 & 28 & 240 \\
$t_s$ & Myr & Time scale & 33 & 3.3 & 2.6 & 180 \\ 
$x_0$ & pc & Injection width & 67 & 67 & 0.1 & 67 \\ \hline
\multicolumn{3}{c}{Dimensionless timescales} \\ \hline
$\tau_f$ & - & Star formation duration & 300 & 3000 & 3000 & 50 \\
$\tau_0$ & - & Injection width & 0.22 & 2.2 & $6\times 10^{-6}$ & 0.039 \\
\hline
\end{tabular}
\caption{
\label{tab:mw_params}
Estimated values of parameters for the Milky Way at the Solar Circle, for various nucleosynthetic sites. SNII and SNIa indicate type II and type Ia supernovae, AGB indicates asymptotic giant branch stars, and NSM stands for neutron star merger. Note that the event rate given for SNII is the rate for clusters of SNe, rather than for individual ones. The dispersion of injection masses for NSM is essentially unconstrained, so we have adopted an arbitrary value that we can use when one is needed for numerical evaluation, with the caveat that any results based on this choice are highly uncertain.
}
\end{table*}

\subsubsection{Diffusion Coefficient}

As noted above, linear diffusion is only a very crude approximation to the process of turbulent transport. At the order of magnitude level, however, the diffusion coefficient associated with turbulence is of order the outer scale of the turbulence multiplied by the turbulent velocity; \citet{karlsson13a} suggest $\kappa \approx \ell \sigma_g/3$, where $\ell$ is the outer scale of the turbulence (typically the ISM scale height), and $\sigma_g$ is the gas turbulent velocity dispersion on this scale. In the Milky Way near the Solar Circle, the neutral ISM has a scale height $h \approx 150$ pc and velocity dispersion $\sigma_g \approx 7$ km s$^{-1}$ \citep{kalberla09a}, so a characteristic Solar neighbourhood value of $\kappa$ is $\sim 300$ pc km s$^{-1}$. More generally, if we have
\begin{equation}
\kappa \approx \frac{h\sigma_g}{3},
\end{equation}
it is convenient to express the scale height as \citep[e.g.,][]{forbes12a}
\begin{equation}
h \approx \frac{f_g \sigma_g^2}{\pi G \Sigma},
\end{equation}
where $\Sigma$ is the gas surface density and $f_g = \Sigma / [\Sigma+(\sigma_g/\sigma_*)\Sigma_*]$ is the effective gas fraction, accounting for the stellar surface density $\Sigma_*$ as modified by differences in the stellar and gas velocity dispersions, $\sigma_g$ and $\sigma_*$. If we eliminate $\Sigma$ by demanding that the gas \citet{toomre64a} parameter
\begin{equation}
Q \approx \frac{\sqrt{2} \sigma_g\Omega}{\pi G \Sigma} \approx 1,
\end{equation}
where $\Omega$ is the angular velocity of the galaxy and the $\sqrt{2}$ factor assumes a flat rotation curve, then we can express the diffusion coefficient as
\begin{equation}
\kappa \approx \frac{f_g Q}{6\sqrt{2} \pi} \sigma_g^2 t_{\rm orb} \approx 190\left(\frac{f_g}{0.5}\right) \sigma_{g,1}^2 t_{\rm orb,2}\mbox{ pc km s}^{-1}
\end{equation}
where $t_{\rm orb}=2\pi/\Omega$ is the galaxy orbital period, $\sigma_{g,1} = \sigma_g/10$ km s$^{-1}$, and $t_{\rm orb,2} = t_{\rm orb}/100$ Myr. In the numerical evaluation we have taken $Q=1$.

\subsubsection{Injection Rate and Mass Variance}
\label{sssec:rate_variance}

The injection rate and injection mass distribution will depend on the primary astrophysical origin site of the element in question, and thus we must distinguish a number of cases. Our treatment of nucleosynthetic origin here will be relatively crude, since our goal is simply to develop order of magnitude estimates for $\Gamma$ and $\sigma_w$.\\

\textit{Type II supernovae} are the main origin sites for $\alpha$ elements. Due to the short lifetimes of the stars that produce them, these events are highly clustered in space and time. Their clustering statistics are complex, as noted above, but as a crude estimate we note that the distributions of both star cluster masses \citep[e.g.,][]{fall12a, bastian12a, adamo17a} and ionising luminosities \citep{kennicutt89a, mckee97a, murray10b} can be fit by a truncated powerlaw with index close to $-2$ or slightly shallower, and a truncation that corresponds to a star clusters mass $M_{\rm cl,max} \sim 10^5$ $M_\odot$. These statements hold across a wide range of galaxy types. For elements produced by type II supernovae, we therefore adopt a cluster mass distribution $dn/dM_{\rm cl}\sim M_{\rm cl}^{-\alpha}$ over the range $M_0$ to $M_1$, and take the event rate equal to the star cluster formation rate: $\Gamma = \dot{\Sigma}_* / \langle M_{\rm cl}\rangle$, where for $q = M_1/M_0 \gg 1$
\begin{equation}
\langle M_{\rm cl} \rangle = \int_{M_0}^{M_1} M_{\rm cl} \frac{dn}{dM_{\rm cl}} dM_{\rm cl}
= M_0 \left\{
\begin{array}{ll}
\frac{\alpha-1}{2-\alpha} q^{2-\alpha}, & 1 < \alpha < 2 \\
\ln q, & \alpha = 2
\end{array}
\right.
\end{equation}
If we take the minimum cluster mass to be $M_0 \approx 100$ $M_\odot$, comparable to the mass of a single massive star, then for $\alpha=2$ and $M_1 \approx 10^5$ $M_\odot$ we have $\langle M_{\rm cl}\rangle \approx 690$ $M_\odot$. The Solar neighbourhood star formation rate is $\dot{\Sigma}_* \approx 2.5\times 10^{-3}$ $M_\odot$ pc$^{-2}$ Myr$^{-1}$ \citep{fuchs09a}, so this implies $\Gamma \approx 3\times 10^{-6}$ pc$^{-2}$ Myr$^{-1}$ near the Sun. More generally, it is convenient to express the star formation rate in terms of the orbital period using the ``dynamical" form of the \citet{kennicutt98a} relation, $\dot{\Sigma}_* = \epsilon_{\rm orb} \Sigma \Omega / 2\pi$, where $\epsilon_{\rm orb} \approx 0.1$ is the fraction of the gas transformed into stars per Galactic rotation. With this substitution, we can express the type II injection rate as
\begin{equation}
\label{eq:gamma_snii}
\Gamma \approx \frac{2\sqrt{2}\epsilon_{\rm orb}}{G Q\langle M_{\rm cl}\rangle} \frac{\sigma_g}{t_{\rm orb}^2} \approx 9.3\times 10^{-5} \frac{\sigma_{g,1}}{t_{\rm orb,2}^2}\mbox{ pc}^{-2}\mbox{ Myr}^{-1},
\end{equation}
where the numerical evaluation is for $\epsilon_{\rm orb} = 0.1$, $\langle M_{\rm cl} \rangle = 690$ $M_\odot$, and $Q=1$. Finally, if the mass injected per cluster is proportional to the cluster mass, then we can integrate over the distribution of cluster masses to obtain
\begin{equation}
\sigma_w^2 = -1 + \left\{
\begin{array}{ll}
\frac{(2-\alpha)^2}{(3-\alpha)(\alpha-1)}\frac{q^{3+\alpha}}{(q^2-q^\alpha)^2}, & 1 < \alpha < 2 \\
q/(\ln q)^2, & \alpha = 2
\end{array}
\right..
\end{equation}
For our fiducial choices, $\alpha = 2$ and $q = 10^3$, we have $\sigma_w^2 \approx 20$; if we instead adopt $\alpha=1.7$, comparable to the shallowest reported indices \citep[e.g.,][]{murray10b} the effect is modest: we find $\sigma_w^2 \approx 15$ instead. In either case, our computed value of $\sigma_w$ probably somewhat underestimates the true dispersion, since it neglects variations in yield within a star cluster of fixed mass as a result of stochastic sampling of the IMF. We will study this effect in future work.\\

\textit{Type Ia supernovae} dominate production of most iron peak elements. They come from older stellar populations that have fully phase-mixed in their host galaxies. Thus we can neglect clustering and instead treat each type Ia supernova as a single event. In the present-day Milky Way, there is $\sim 0.01$ yr$^{-1}$ type Ia supernova per year, spread out over the $\approx 30$ kpc$^2$ effective area of the Galactic stellar disc, giving $\Gamma \sim 3\times 10^{-4}$ pc$^{-2}$ Myr$^{-1}$. In general this rate will be set by the convolution of the star formation history with the (poorly-constrained) type Ia supernova delay time distribution. The spread in mass injected is considerably smaller than for type II SNe due to the lack of clustering. \citet{scalzo14a} show that the dispersion in $^{56}$Ni mass is relatively narrow, with the great majority of the observed SN Ia producing $0.3 - 0.6$ $M_\odot$ of $^{56}$Ni. Examining their Figure 1 suggests $\sigma_w^2 \sim 0.1$. Theoretical models also predict relatively modest spreads for most iron peak elements, with some possible exceptions (e.g., Mn -- \citealt{seitenzahl13a}; see the review by \citet{seitenzahl17a} for more discussion). We will adopt $\sigma_w^2 = 0.1$ as a fiducial value. \\

\textit{AGB stars} dominate production of $s$-process elements. While the stars that produce the $s$-process are not as old as those that produce type Ia supernovae, for most elements the dominant production sites are stars smaller than $\sim 5$ $M_\odot$ \citep[e.g.,][]{karakas16a}, which have lifetimes $> 100$ Myr. This is much longer than the typical star remains clustered together with those born nearby. It is therefore reasonable to treat these stars as phase-mixed as well. Thus the event rate is simply the formation rate of stars that end their lives in the AGB phase. For a \citet{chabrier05a} IMF, one star in the mass range $1-8$ $M_\odot$ form per $\sim 5$ $M_\odot$ of stars formed, so the event rate in a region with a star formation rate per unit area $\dot{\Sigma}_*$ is $\Gamma \approx \dot{\Sigma}_*/(5\, M_\odot)$. Using the Solar neighbourhood star formation rate, this implies $\Gamma \approx 5\times 10^{-4}$ pc$^{-2}$ Myr$^{-1}$, and for the more general case the AGB star injection rate will be given by \autoref{eq:gamma_snii} evaluated with a ``cluster mass" $\langle M_{\rm cl}\rangle = 5$ $M_\odot$, and will therefore be $\approx 140$ times the type II supernova rate. The dispersion varies from element to element, depending on exactly where in the AGB mass range dominates production, but is typically a factor of a few; for numerical purposes we will adopt $\sigma_w^2 = 5$ as a typical value. \\
 
\textit{Neutron star mergers (NSM)} may be the dominant site of the $r$-process, though this remains highly uncertain. Estimates suggest that the present-day Milky Way experiences $\sim 10$ such mergers per Myr \citep[e.g.,][]{van-de-voort15a, shen15a}; these are likely spread over a somewhat larger area than the stars, due to asymmetric supernova kicks. We therefore estimate an effective rate $\Gamma \sim 10^{-7}$ pc$^{-2}$ Myr$^{-1}$. The distribution of mass injected (as opposed to the mean) is at present completely unconstrained. For lack of a better choice we will adopt $\sigma_w^2 = 10$ for these events when a value is necessary for numerical evaluation, but one should keep in mind that this value is uncertain by at least an order of magnitude.

\subsubsection{Length and Time Scales}

We are now in a position to calculate the characteristic length and time scales $x_s$ and $t_s$ (\autoref{eq:xs} and \autoref{eq:ts}). For conditions near the Solar Circle in the Milky Way, using our stated estimates of $\kappa$ and $\Gamma$, we have
\begin{equation}
\left(\frac{x_s}{\mbox{pc}}, \frac{t_s}{\mbox{Myr}}\right) = \left\{
\begin{array}{ll}
(100, 33), & \mbox{SNII} \\
(32, 3.3), & \mbox{SNIa} \\
(28, 2.6), & \mbox{AGB} \\
(240, 180), & \mbox{NSM}
\end{array}
\right.
\end{equation}
for the various potential astrophysical origin sites.

Using our more general expressions for SNII and AGB stars, which are linked directly to the star formation rate, we have
\begin{eqnarray}
x_s & \approx & \left(\frac{f_g G Q^2 \langle M_{\rm cl}\rangle}{24\pi \epsilon_{\rm orb}}\sigma_g t_{\rm orb}^3 \right)^{1/4} 
\nonumber \\
& \approx & (38,11) \left(\frac{f_g}{0.5}\right)^{1/4} \sigma_{g,1}^{1/4} t_{\rm orb,2}^{3/4} \mbox{ pc} \\
t_s & \approx & \left(\frac{3\pi G \langle M_{\rm cl}\rangle}{\epsilon_{\rm orb}} \frac{t_{\rm orb}}{f_g \sigma_g^3}\right)^{1/2}
\nonumber \\
& \approx & (7.4, 0.63) \left(\frac{f_g}{0.5}\right)^{-1/2} \sigma_{g,1}^{-3/2} t_{\rm orb,2}^{1/2}\mbox{ Myr}
\label{eq:ts_num}
\end{eqnarray}
where the first coefficient in parentheses is for type II SN and the second is for AGB stars. The numerical evaluations use our fiducial values for all parameters.

\subsubsection{Dimensionless Time Scales}

The next step in applying our formalism to astrophysical systems is to estimate the two dimensionless parameters $\tau_f$ and $\tau_0$ that enter our results. The first of these is relatively straightforward: $\tau_f$ is simply the time over which stars formed in the system in question, measured in units of $t_s$. For the Milky Way disc the formation time is $\approx 10$ Gyr, so for type II supernova species $\tau_f \approx 300$; the corresponding figure for type Ia supernovae and AGB stars is $\tau_f \sim 3000$, while for neutron star mergers $\tau_f \sim 50$. For our more general expression for type II supernovae and AGB stars, we have (\autoref{eq:ts_num}),
\begin{eqnarray}
\tau_f & \approx & \left(\frac{\epsilon_{\rm orb}}{3 \pi G \langle M_{\rm cl}\rangle} \frac{f_g \sigma_g^3 t_*^2}{t_{\rm orb}}\right)^{1/2}
\nonumber \\
& = & \left(1.4 \times 10^3, 1.6\times 10^4\right) \left(\frac{f_g}{0.5}\right)^{1/2} \sigma_{g,1}^{3/2} t_{\rm orb,2}^{-1/2} t_{*,1},
\end{eqnarray}
where $t_{*,1}$ is the age of the system in units of 10 Gyr and all parameters have their fiducial values.

The dimensionless time scale $\tau_0$, which parameterises the ``initial" radius over which injected metals are dispersed, is slightly more uncertain, though this uncertainty is mitigated by the fact that the results are only logarithmically sensitive to it in most cases. We can think of this quantity as describing the mass of ISM swept up by metal-rich ejecta before they halt systematic expansion and begin to be diffused by the turbulence in the ISM. Formally, $\tau_0 = \sigma_{\rm inj}^2/2$, where $\sigma_{\rm inj}$ is the characteristic dimensionless radius at which this transition occurs.

For supernovae of any type, and for neutron star merger remnants (which behave similarly in their interactions with the ISM -- \citealt{montes16a}), \citet{draine11a} shows (his equation 39.31) that the characteristic radius at which the blast wave expansion velocity becomes equal to the ISM velocity dispersion $\sigma_g$ is
\begin{equation}
x_0 \approx 67 E_{51}^{0.32} n_0^{-0.37} \sigma_{g,1}^{-2/5}\mbox{ pc},
\end{equation}
where $E_{51}$ is the supernova energy in units of $10^{51}$ erg and $n_0$ is the ISM number density in units of hydrogen nuclei per cm$^3$. If we take $x_0/x_s$ as a rough estimate of $\sigma_{\rm inj}$, then for the Milky Way (which has $n_0 \approx 1$) this gives $\tau_0 \approx 0.22$, 2.2, or 0.039 for type II supernovae, type Ia supernovae, and neutron star mergers, respectively. For the purposes of more general evaluation in the type II supernova case, it is convenient to re-express the ISM density in terms of other parameters; \citet[their equation 23]{krumholz17c} show that the midplane density is $\rho \approx 2 \Omega^2/\pi G Q^2 \phi_{\rm mp} f_g$, where $\phi_{\rm mp}\approx \sqrt{2}$ is the ratio of total to turbulent pressure. Using this expression to evaluate $n_0$, and using the value of $x_s$ for type II supernovae, gives
\begin{equation}
\tau_0 \approx 0.15 \left(\frac{f_g}{0.5}\right)^{0.24} \sigma_{g,1}^{-13/10} t_{\rm orb,2}^{-0.02}.
\end{equation}

The ejecta produced by AGB stars have much lower velocity, and as a result they expand much less before halting and mixing. Observationally, the typical radii of AGB star bubbles are typically $\sim 0.1$ pc, though with a wide range of variation \citep[e.g.][]{cox12a}. Adopting this as a fiducial estimate of $x_0$, we have $\tau_0 \sim 6\times 10^{-6}$ for AGB stars near the Solar Circle.

\subsection{Metallicity Dispersions}

Having worked out the physical and dimensionless scales in our problem, we now present our first application: computing the metallicity dispersion we might expect to find in the ISM, or in a population of young stars. The usual measure of variation in observed metallicities is the variance in the logarithm (base 10) of metallicity, i.e., for a given set of measured metallicities $Z_i$ (in gas or in stars), the quantity of interest is
\begin{equation}
\sigma_{\log Z}^2 \equiv \left\langle\left(\log Z_i  - \langle\log Z_i\rangle\right)^2\right\rangle,
\end{equation}
where the angle brackets indicate averaging over the sample. When the range of variation in $Z_i$ about the mean measurement $\langle Z\rangle$ is small, which is the case for all astrophysically-relevant cases, we can Taylor expand the logarithms about $Z_i / \langle Z\rangle = 1$, which gives
\begin{equation}
\sigma_{\log Z}^2 \approx \left(\frac{1}{\langle Z \rangle\ln 10 }\right)^2 \left\langle\left(Z_i - \langle Z\rangle\right)^2\right\rangle.
\end{equation}
For a constant gas surface density, we can rewrite this in terms of our dimensionless quantities as
\begin{equation}
\sigma_{\log Z}^2 \approx \left(\frac{1}{\langle S_X\rangle\ln 10}\right)^2 \sigma^2
= \frac{1+\sigma_w^2}{8\pi \tau_f (\ln 10)^2} \ln \left(1 + \frac{\tau_f}{\tau_0}\right).
\end{equation}

We provide one caveat on this expression, which is that our simple analytic model does not include galactic winds, which will remove metals over time. To see how to include these, we must consider two possibilities. First suppose that the winds simply remove portions of the ISM that are widely-distributed and uncorrelated with the metal field. These removed regions simply carry with them whatever metals they contained at the time of removal. In this limit the effect of metal removal on the dispersion of the metal field is small, and the sole effect of removal is to lower the mean metal content; in terms of our dimensionless solution, in this limit winds do not alter $\sigma$, but they reduce $\langle S_X\rangle$ by a factor $f_d$, where $f_d$ is the fraction of the metal retained in the disc rather than lost to the wind. This in turn will raise $\sigma_{\log Z}$. This is the case that is likely to prevail for elements injected by AGB stars, type Ia supernovae, and neutron star mergers, since none of these are capable of launching galactic winds. For type II supernovae, on the other hand, the situation may be different, depending on whether the metals ejected are primarily those that were mixed with the ISM before the supernova explosions, or are primarily unmixed supernova ejecta. In the former case the situation is the same as with other nucleosynthetic sites. In the latter case, on the other hand, the effect is simply to reduce the mass of metals $m_X$ injected by each event. Since $m_X$ does not affect $\sigma/\langle S_X\rangle$, in this case winds do not alter $\sigma_{\log Z}$ at all.

Combining these two cases, we empirically modify our expression for the metallicity dispersion to
\begin{equation}
\sigma_{\log Z}^2 \approx \frac{1+\sigma_w^2}{8\pi f_d^2 \tau_f(\ln 10)^2} \ln \left(1 + \frac{\tau_f}{\tau_0}\right),
\label{eq:sigma_logZ}
\end{equation}
where $f_d$ is a factor equal to the fraction of metals retained in the disc for elements whose primary origin site is not type II SNe, and is a factor between that and unity for elements that do come primarily from type II SNe. In the Milky Way and similar galaxies, the observed fraction of disc metals is $f_d \approx 0.5$ \citep[e.g.,][]{tumlinson11a, werk14a}. Inserting our fiducial Milky Way values from \autoref{tab:mw_params}, we obtain
\begin{equation}
\sigma_{\log Z} \approx 
\left\{
\begin{array}{ll}
0.12, &\mbox{(SNII)} \\
0.009 & \mbox{(SNIa)} \\
0.035, & \mbox{(AGB)} \\
0.22, & \mbox{(NSM)}
\end{array}
\right.,
\end{equation}
where we have adopted $f_d = 0.5$ for SNII, recalling that the true value could be closer to unity, which in turn would decrease $\sigma_{\log Z}$ as $f_d^{-1}$; this introduces an uncertainty of somewhat less than a factor of 2. Also note that our estimate for NSM is based on our rather arbitrarily-chosen value of $\sigma_w$ for these events, and thus should be regarded is extremely uncertain.

Using our more general expressions for $\tau_f$ and $\tau_0$ and their relationship to the star formation rate in the case of SNII-borne elements, we have
\begin{eqnarray}
\sigma_{\log Z} & \approx &
0.065 \left(\frac{f_d}{0.5}\right)^{-1} \left(\frac{f_g}{0.5}\right)^{-1/4} \sigma_{g,1}^{-3/4} t_{\rm orb,2}^{1/4} t_{*,1}^{-1/2} \cdot {}
\nonumber \\
& &
\left(1 + 0.029 \ln \frac{f_g}{0.5} + 0.31 \ln \sigma_{g,1} 
\right.
\nonumber \\
& &
\left.
\quad 
{}-0.053\ln t_{\rm orb,2} + 0.11 \ln t_{*,1}
\phantom{\frac{f_g}{0.5}\hspace{-0.2in}}
\right)^{1/2}
\label{eq:sigma_logz_sn}
\end{eqnarray}
where we have approximated $\tau_f/\tau_0 \gg 1$. The equivalent expression for AGB-produced elements, using the same value of $\tau_0$ everywhere for lack of an observational basis on which to vary this value, is
\begin{eqnarray}
\sigma_{\log Z} & \approx & 0.016 \left(\frac{f_d}{0.5}\right)^{-1} \left(\frac{f_g}{0.5}\right)^{-1/4} \sigma_{g,1}^{-3/4} t_{\rm orb,2}^{1/4} t_{*,1}^{-1/2} \cdot {}
\nonumber \\
& &
\left(1 + 0.023 \ln \frac{f_g}{0.5} + 0.069 \ln \sigma_{g,1}
\right.
\nonumber \\
& &
\left.
\quad 
{}-0.023\ln t_{\rm orb,2} + 0.046\ln t_{*,1} 
\phantom{\frac{f_g}{0.5}\hspace{-0.2in}}
\right)^{1/2}
\end{eqnarray}
Thus we arrive at a robust first-principles explanation for why the typical metallicity dispersion for $\alpha$ elements in nearby disc galaxies is $\sim 0.1$.

\subsection{Spatial Correlations}

We next examine the correlations in space predicted by our model. Note that, although we have computed the spatial correlation of the metal field, because there are no gas dynamics in our simple model we are implicitly assuming a constant gas density. Since movements of gas will change the metal density but not the metallicity, we should therefore think of our prediction of $\xi$ as being a prediction of the spatial correlation of metallicity rather than metal density. We also note that, in the case of small variations, the correlation is the same whether we consider absolute or logarithmic metallicities.

Before proceeding to a numerical evaluation, it is helpful to consider how we expect the correlation at fixed physical (as opposed to dimensionless) scale to behave as we alter dimensional quantities. Consider a region with diffusion coefficient $\kappa$ and injection rate $\Gamma$, where star formation has been going on for a physical time $t_*$. Moreover, suppose that injected metals are dispersed over an initial physical radius $x_0$, which is related to $\tau_0$ by $\tau_0 = (x_0/x_s)^2/2$. Writing \autoref{eq:xi_final} in terms of the physical variables $x = r x_s$ and $t_* = \tau_f t_s$, and making a change of variables $a = (\Gamma/\kappa)^{1/4} k$ in the integral, the correlation becomes is
\begin{equation}
\xi(x) = \frac{2}{\ln\left(1+\frac{2 \kappa t_*}{x_0^2}\right)}
\int_0^\infty e^{-x_0^2 a^2} \left(1-e^{-2 \kappa t_* a^2}\right)
\frac{J_0\left(a x\right)}{a} \, da.
\label{eq:xi_phys}
\end{equation}
The important point to take from this expression is that the correlation at a given fixed physical length $x$ is independent of the injection rate $\Gamma$ or the dispersion in mass injected $\sigma_w$. It depends only on the physical injection radius $x_0$ and the product of the time over which metal injection has taken place $t_*$ and the diffusion coefficient $\kappa$. Of these quantities, only $x_0$ could plausibly be different between different nucleosynthetic sites, and we expect $x_0$ to be nearly the same for type Ia SN, type II SN, and neutron star mergers, since all three launch blast waves with comparable energy budgets; only AGB stars will differ. Thus even without performing any numerical evaluations, we arrive at the interesting conclusion that elements whose dominant formation sites are either type of supernova or neutron star mergers should all show nearly the same spatial correlation; only elements that originate in AGB stars should appear different.

\begin{figure}
\includegraphics[width=\columnwidth]{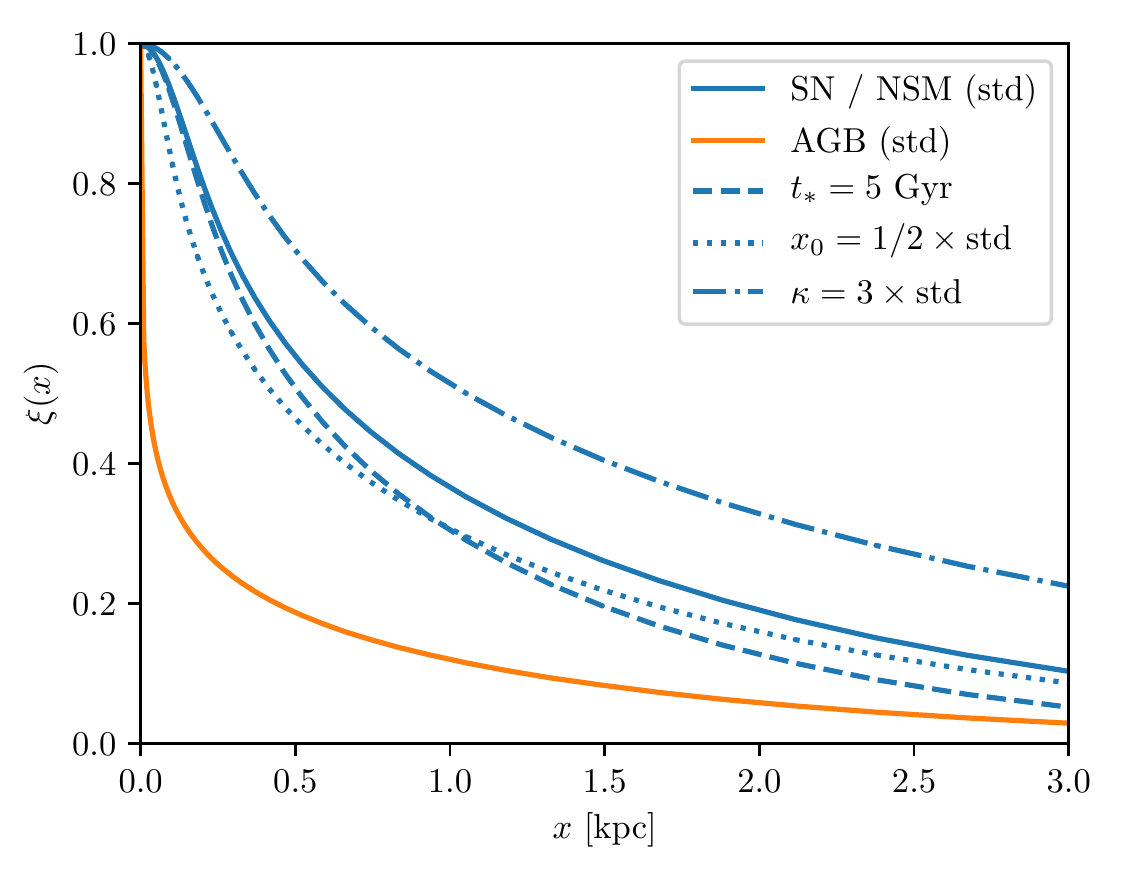}
\caption{
\label{fig:corr_mw}
Pearson correlation of metals $\xi(x)$ as a function of separation $x$, computed for Milky Way Solar Circle parameters (solid lines, \autoref{tab:mw_params}). The two solid lines use our standard parameters; one shows results for elements whose origin sites are supernovae or neutron star mergers (which all have the same correlation -- see main text) and one shows elements produced by AGB stars. Various other line styles show results for supernovae and neutron star mergers using variations on our standard parameter choices: a star formation duration of 5 Gyr rather than 10 Gyr (dashed), half the standard supernova blast wave merger radius (dotted), and three times the standard diffusion coefficient (dot-dashed).
}
\end{figure}

We show the correlation at constant time computed for our fiducial Milky Way parameters (\autoref{tab:mw_params}) in \autoref{fig:corr_mw}. First focus on the solid lines, which show our standard parameter choices. One interesting conclusion to draw from this figure is that, for elements produced in supernovae or mergers, there are non-negligible correlations in abundances even over kpc scales. The correlation does not drop to $0.5$ until the separation reaches 600 pc, and remains above 30\% out to distance of $1.3$ kpc. Elements produced by AGB stars are substantially less correlated, with significant correlations confined to $\lesssim 100$ pc scales.

Now consider the other line styles, which show the sensitivity of our results to various parameter choices. We see that the correlation is relatively insensitive to $x_0$. It depends in the same way on $\kappa$ and $t_*$ (since these enter as a product), but since in real systems $\kappa$ is the more uncertain of the two parameters, we can regard it as the one most likely to affect the results. The effect of varying $\kappa$ can be approximated well by simply rescaling the correlation function by a factor of $\kappa^{1/2}$, i.e., increasing the diffusion coefficient by a factor of 3 causes the physical scale that corresponds to a particular correlation value to increase by a factor of nearly $\sqrt{3}$.

\begin{figure}
\includegraphics[width=\columnwidth]{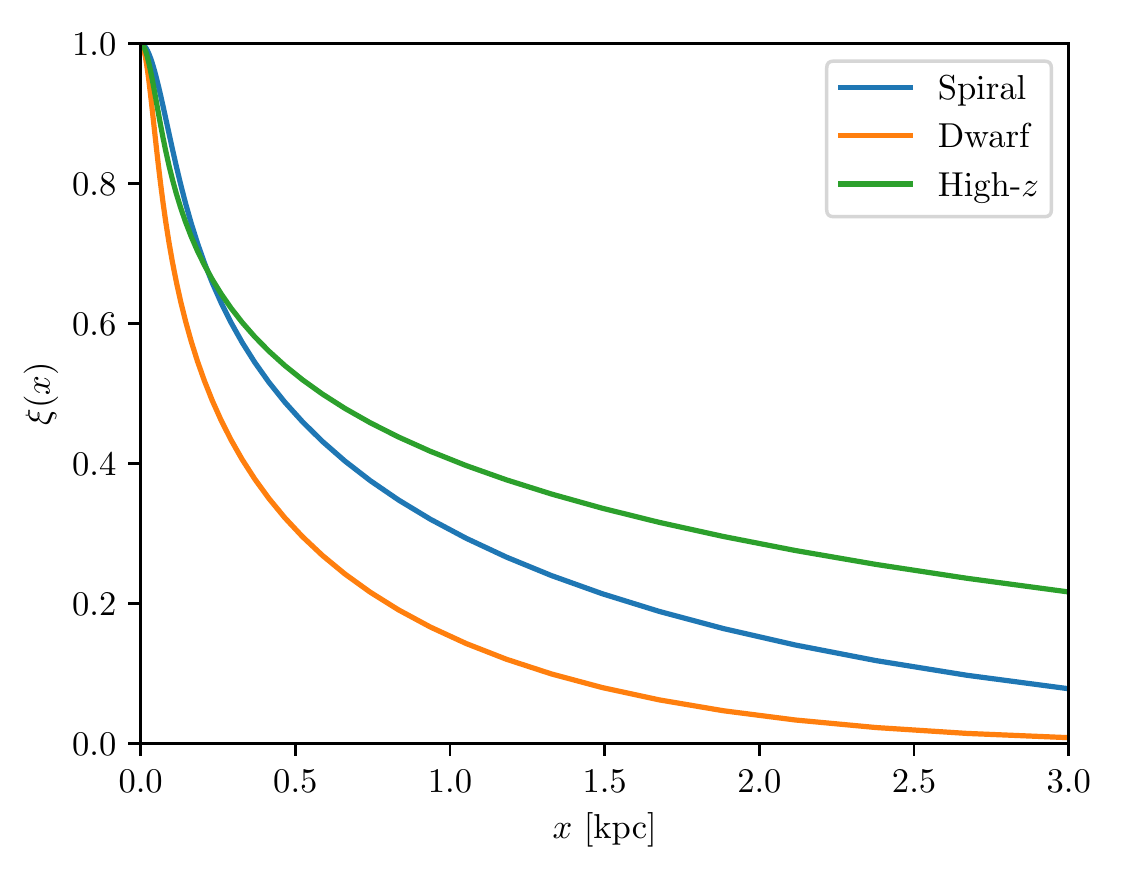}
\caption{
\label{fig:corr_exgal}
Pearson correlation for metals produced by supernova-like injection events, evaluated for parameters appropriate to local spiral galaxies, local dwarf galaxies, and $z\approx 2$ star-forming discs. The exact parameters used are as follows: $f_g = (0.5, 0.9, 0.7)$, $\sigma_{g,1} = (0.8, 0.6, 4)$, $t_{\rm orb,2} = (2, 0.5, 2)$, and $t_{*,1} = (1, 1, 0.25)$. Here the first entry is the local spiral case, the second is the local dwarf case, and the third is the high-$z$ case.
}
\end{figure}

We can also use \autoref{eq:xi_phys} to investigate how the correlation scale should change with galactic properties. Examining the expression, we see that the physical correlation length is sensitive to only two physical length scales, $x_0$ and $\sqrt{2 \kappa t_*}$. The former is an effective radius over which a single event injects metals, and we can think of the latter as the effective length scale over which diffusion spreads metals in a time $t_*$. Using the physical scalings established above, and considering the case of supernova-like injection events, we expect these to vary with macroscopic galactic properties as
\begin{eqnarray}
x_0 & \approx & 21 \left(\frac{f_g}{0.5}\right)^{0.37} \sigma_{g,1}^{-2/5} t_{\rm orb,2}^{0.74}\mbox{ pc} \\
\sqrt{2 \kappa t_*} & \approx & 2.0 \left(\frac{f_g}{0.5}\right)^{1/2} \sigma_{g,1} t_{\rm orb,2}^{1/2} t_{*,1}^{1/2}\mbox{ kpc},
\end{eqnarray}
where we have taken $E_{51} = Q = 1$ in the numerical evaluation.

It is interesting to use these scalings to evaluate the correlation for parameters typical of different types of galaxies. We do so in \autoref{fig:corr_exgal}, using parameter choices appropriate for local spirals, local dwarfs, and $z\approx 2$ star-forming discs. The plot shows that, at a fixed physical scale, we expect the metallicities in high-$z$ discs to be the most correlated and local dwarfs to be the least correlated, with local spirals in between.

\begin{figure}
\includegraphics[width=\columnwidth]{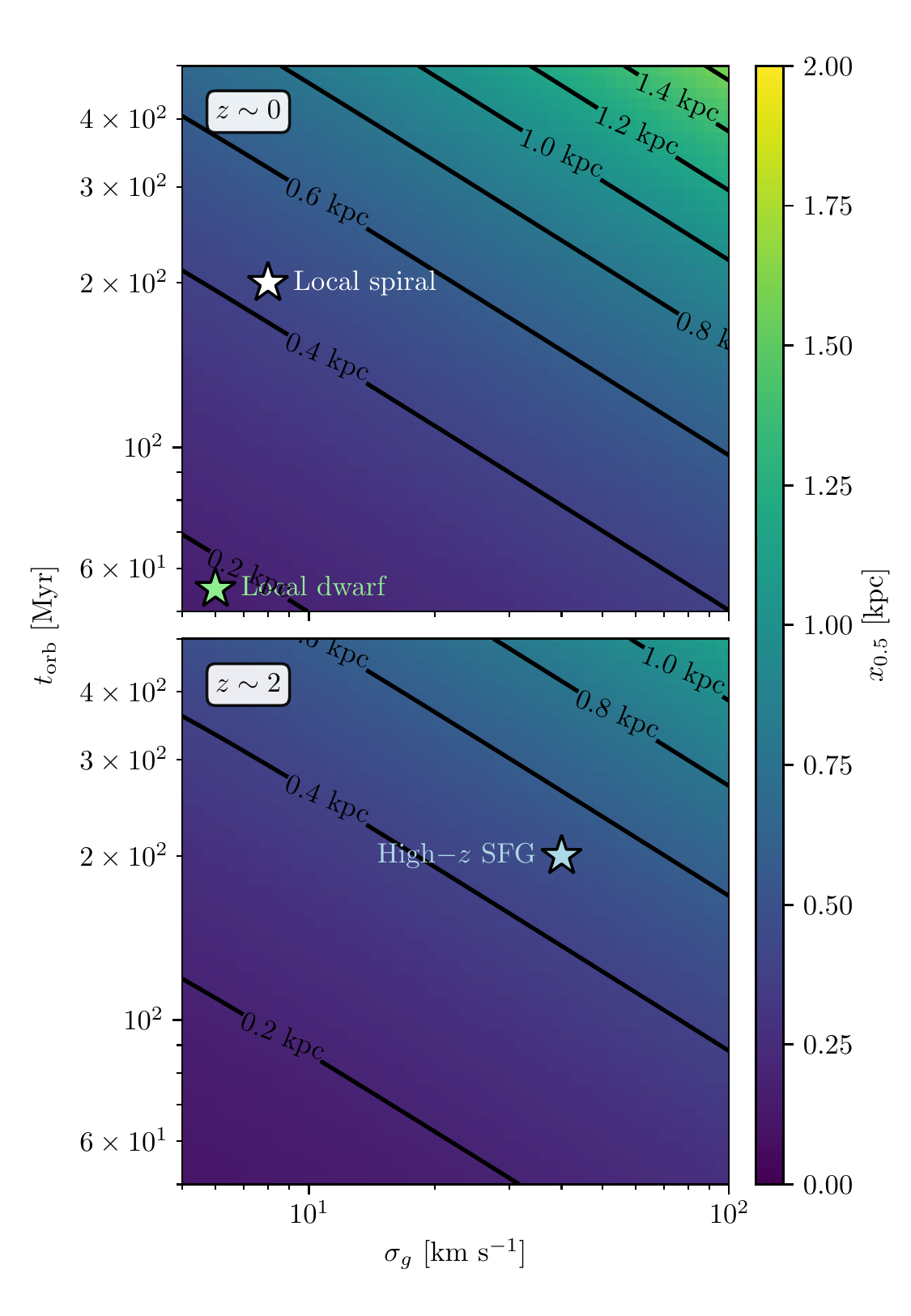}
\caption{
\label{fig:corr_exgal_contour}
Correlation length $x_{0.5}$, defined as the length scale for which the Pearson correlation reaches 0.5 for type II supernova-produced elements, as a function of ISM velocity dispersion $\sigma_g$ and galaxy orbital period $t_{\rm orb}$. The top panel shows the results for galaxies with a star formation age $t_*  = 10$ Gyr, appropriate for the present day, while the bottom shows galaxies at $t_* = 2.5$ Gyr, appropriate for $z\sim 2$. All calculations use $f_g = 0.5$. Labels stars indicate the approximate locations of the ``typical" local spiral, local dwarf, and high$-z$ star-forming galaxy parameters shown in \autoref{fig:corr_exgal}. 
}
\end{figure}

We summarise the dependence of the correlation on galaxy properties in \autoref{fig:corr_exgal_contour}. In this figure we show contour plots of $x_{0.5}$, defined by the implicit relation $\xi(x_{0.5}) = 0.5$, i.e., $x_{0.5}$ is the length scale for which the correlation of the metal field is 50\%. The figure shows two different star formation ages $t_*$, one appropriate to modern galaxies, and one appropriate to systems at $z\sim 2$. For a fixed system age, the correlation length clearly increases toward higher velocity dispersion and higher orbital periods. Thus in general we expect that smaller galaxies, which tend to have smaller orbital periods and velocity dispersions, will have less highly-correlated metal fields than larger galaxies. If one held all other galaxy properties fixed, the correlation would grow with cosmic time. However, the mean velocity dispersions of galaxies decreases with redshift \citep[e.g.,][]{wuyts16a}, and the gas fraction increases with redshift \citep[e.g.,][]{tacconi13a}. This compound effect pushes high redshift galaxies to higher correlation lengths; therefore, at least at fixed stellar or halo mass, galaxies tend to have a longer correlation length at higher redshift.

\subsection{Space and Time Correlations}

Making the same substitution from dimensionless to physical variables as in the previous section, the space-time correlation function (\autoref{eq:xi_t_final}) is
\begin{eqnarray}
\lefteqn{
\xi(x,\Delta t) = \frac{2}{\sqrt{
\ln\left(1+\frac{2 \kappa t_*}{x_0^2}\right) \ln\left[1 + \frac{2\kappa (t_*-\Delta t)}{x_0^2}\right]}}
\cdot {}
}
\nonumber \\
& &
\int_0^\infty e^{-(\kappa \Delta t+x_0^2)a^2}
\left[1 - e^{-2\kappa(t_*-\Delta t)a^2}\right]
\frac{J_0(a x)}{a} \, da,
\end{eqnarray}
where $\Delta t$ is the separation in physical time. As with the correlation at fixed time, $\Gamma$ and $\sigma_w$ do not enter, and thus both the correlation in time and the correlation in space are expected to be the same for all astrophysical origin sites that inject their products over comparable physical size scales in the ISM. 

\begin{figure}
\includegraphics[width=\columnwidth]{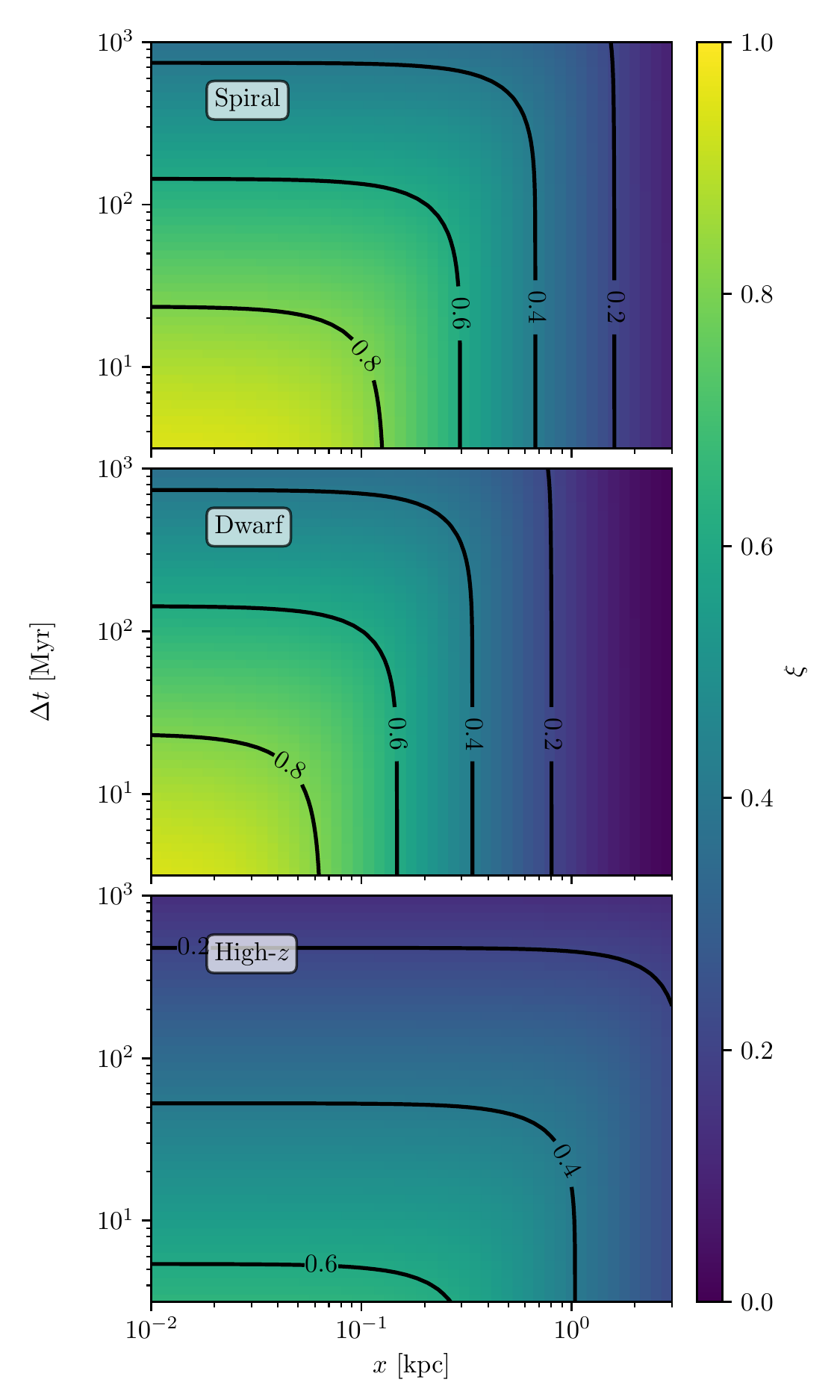}
\caption{
\label{fig:corr_time}
Pearson correlation $\xi$ of the metal field produced by supernova- or supernova-like injection events as a function of separation in space $x$ and time $\Delta t$. The three panels show three sets of example parameters, appropriate for local spirals, local dwarfs, and high-$z$ star-forming galaxies, as indicated. Parameter values are the same as those used in \autoref{fig:corr_exgal}.
}
\end{figure}

In \autoref{fig:corr_time} we show the predicted correlation as a function of space and time separation for elements injected by supernova-like explosions, for the same three example parameter sets (local spirals, local dwarfs, and high-$z$ star-forming galaxies) as in \autoref{fig:corr_exgal}. We see that local galaxies show significant correlations in their metal fields at fixed position even over timescales of $\sim 100$ Myr. The correlation does not drop below $\xi = 0.5$ until times of $\sim 300$ Myr in either local dwarfs or spirals. Note that, because we have not included rotation in our simple model, this should be interpreted as the correlation at a fixed Lagrangian position, i.e., at a point orbiting with the galaxy. Nonetheless, this result implies that ISM element abundances in local spirals take a relatively long time to ``forget" their prior states and re-randomise. At high redshift, on the other hand, this shuffling occurs much more quickly, with 50\% correlation loss on timescales of only 10 Myr. The primary reason for this difference is that, at small spatial separations, the correlation is determined by $t_* - \Delta t$, and $t_*$ is substantially smaller at $z\sim 2$ than today.

\green{A final implication of the timescales we have derived is that, while events such as mergers or accretion of blobs of low-metallicity gas are important for setting galaxies' overall metallicities and metallicity gradients, they are relatively unimportant for fluctuation statistics. This is simply because, for most galaxies, mergers or significant fluctuations in the rate of accretion from the cosmic web occur over timescales much larger than the correlation time. Thus a merger or similar event that scrambles the metallicity will be ``forgotten" long before the next one occurs.}

\section{Discussion and Conclusions}
\label{sec:conclusion}

In this first paper in a series, we present a formalism for studying metallicity fluctuations in galaxies' interstellar media and young stellar populations. Our formalism treats metallicity evolution as a stochastically-forced diffusive system. While this is obviously a substantial simplification of the true complexity of element injection and transport through the ISM, it captures the central qualitative feature that metallicity statistics result from a competition between stochastic injection events, which produce inhomogeneity, and mixing by interstellar turbulence, which homogenises the gas. Moreover, this system has the virtue that it is simple enough that we can obtain exact analytic solutions for important statistical properties of the metal field and their dependence on the host galaxy and the astrophysical origin site of the element in question.

The major findings of our investigation are as follows:
\begin{itemize}
\item Under the conditions found in the Milky Way near the Solar Circle, the equilibrium dispersion in ISM and young stellar abundance at any given time is expected to range from a low of $\approx 0.01$ dex for elements injected by type Ia supernovae to a high of $\approx 0.2$ dex for elements produced by neutron star mergers. Elements whose origin sites are AGB stars or type II supernovae are intermediate between these two limits, at $\approx 0.04$ and $\approx 0.1$ dex, respectively. Since type II supernova-produced elements are the most easily measured in the gas phase (e.g., O, N), this means that most gas phase abundance measurements in nearby galaxies like the Milky Way should return abundances spreads of $\approx 0.1$ dex. This value is consistent with recent measurements of the oxygen abundance scatter in the ISM of the Milky Way and similar nearby galaxies \citep[e.g.,][]{balser15a, berg15a, vogt17a}, \green{
and with the results of high-resolution non-cosmological simulations \citet{kubryk13a}.}
\item We predict that, for any element, the scatter should vary systematically with galaxy properties as described in \autoref{eq:sigma_logz_sn}. At fixed age, the most important factors in determining this variation are the fraction of metals produced that are retained in the disc, $f_d$ and the gas velocity dispersion $\sigma_g$. Increases in either of these factors reduce the abundance scatter. The age of the system, the gas fraction and the orbital period also play a lesser role. As a result of these dependencies, dwarf galaxies, which systematically retain fewer metals and have (slightly) lower velocity dispersions than spirals, should show larger abundance scatters than spirals at fixed galactocentric radius.
\item This dispersion should show significant spatial correlations. For conditions similar to those at the Milky Way Solar Circle, metal abundances should be 50\% correlated on size scales of $\sim 0.5$ kpc, with correlations at the $20 - 30$\% level persisting out to distances of $\sim 1-2$ kpc. The correlation length will be substantially smaller in dwarf galaxies, $\sim 0.1-0.2$ kpc. These predictions are directly testable using IFU surveys of nearby galaxies, such as SAMI \citep{allen15a}, CALIFA \citep{sanchez12a}, and MaNGA \citep{bundy15a}.
\item Abundances are also correlated in time for surprisingly long periods, implying that stars born at similar locations at two different times will have correlated abundances. The time required for the ISM of the Milky Way to ``forget" its abundances, meaning that the correlation drops below $\sim 50\%$, is $\sim 300$ Myr, implying that ISM abundances at a given position are randomised only about once per orbit. This resetting time is comparable for local dwarfs, but the ISM memory time is much smaller in the high-redshift universe as a result of the lower overall age of galaxies.
\end{itemize}

\green{Correlations in metal fields represent both a challenge and an opportunity for chemical tagging studies.} The challenge is that, for much chemical tagging work, the implicit assumption has been that stars form in discrete clusters that are internally chemically uniform or close to it, but that are essentially uncorrelated from one cluster to another \citep[e.g.,][]{bland-hawthorn10a,ting15a}. This would make it easy to identify and separate clusters. However, if abundances are correlated on scales of kpc and times of hundreds of Myr, the number of ``unique" chemical signatures may be much smaller than had previously been assumed, even if there is sufficient spread in chemical space. Rather than chemical space resolving into discrete and well-separated clusters, stars may populate it in a much more continuous, fractal distribution, exactly as is observed to be the case in modern measurements of the spatial positions and age distributions of young stars. This may thwart some approaches to analysis that rely on cluster-finding in chemical space \citep[e.g.,][]{ting16a, bland-hawthorn16a}, at least in the most stringent context of chemical tagging, where we look for stars that were born in the same cluster instead of an association of star forming regions. However, it opens up new possibilities as well, since we show in this paper that complex structures in chemical space can be mapped onto structures in physical space and time, not just in instantaneous bursts over small spatial scales, but covering a very wide range of space and time scales.

The model we have developed here is very simple. We have not examined the correlation between elements with similar origin sites, which limits our ability to predict full correlations in chemical space. For example, we would expect $\alpha$ element abundances to be correlated with one another, since the same supernovae will produce contribute across multiple elements. Exploring statistics of this sort, and their ability to break degeneracies induced by chemical space correlation, is the subject of the next paper in this series.

We also have not accounted for the radial structure of galaxies, nor for large-scale inhomogeneities such as spiral arms. \green{Radial metallicity gradients, and conservation of angular momentum more generally, will likely make metallicity spatial statistics anisotropic, with different diffusion rates and correlation lengths in the radial and azimuthal directions. Because we have not treated this effect explicitly, in galaxies with strong gradients our model is likely to provide more reliable results for the azimuthal than the radial correlation. In galaxies with strong spiral patterns, arms will also likely imprint features on metallicity statistics at the Toomre scale, the interarm-spacing, or both. Measuring these effects quantitatively} likely requires a campaign of numerical simulations. The correlation length scales we derive in this analysis suggest that these simulations will need to have $\sim 10$ pc or better resolution, so that mixing is not dominated by numerical diffusion on the $\sim 100$ pc natural size scales of the metal field. This requirement suggests these will likely have to be non-cosmological simulations for the time being. Such a program will be required to make sense of the chemical-space data that are already in hand, and the much larger data set that will become available in the next few years.

\section*{Acknowledgements}

MRK is supported by the Australian Research Council Discovery Projects funding scheme, project DP160100695. YST is supported by the Australian Research Council Discovery Project funding scheme, project DP160103747, the Carnegie-Princeton Fellowship 
and the Martin A. and Helen Chooljian Membership from the Institute for Advanced Study in Princeton.

\bibliographystyle{mn2e}
\bibliography{refs}

\begin{appendix}

\section{Proof of Validity of Approximation for $S_X$}
\label{app:proof}

In this appendix we demonstrate that
\begin{eqnarray}
\lefteqn{\lim_{R'\rightarrow \infty} \frac{1}{A} \int_A \left\langle S_X(\mathbf{r}+\mathbf{r}') S_X(\mathbf{r}') \right\rangle \, d^2 r'}
\nonumber \\
& = & 
\lim_{R'\rightarrow \infty}
\frac{1}{A} \int_A \left\langle S_{X,A}(\mathbf{r}+\mathbf{r}') S_{X,A}(\mathbf{r}') \right\rangle \, d^2 r',
\label{eq:SXA_int}
\end{eqnarray}
where $S_{X,A}$ is as defined in \autoref{eq:SXA}, i.e., it includes only those events from $S_X$ that are injected radii $r' < R'$, so that they lie inside $A$. To demonstrate this, let us divide the sum on the left-hand side of \autoref{eq:SXA_int} into event pairs that are both inside $R'$ and pairs where one or both are outside $R'$:
\begin{eqnarray}
\lefteqn{
\left\langle \frac{1}{A} \int_A S_X(\mathbf{r}+\mathbf{r}') S_X(\mathbf{r}') \, d^2 r' \right\rangle
}
\nonumber \\
& = & 
\left\langle \frac{1}{A} \int_A
\left[
S_{X,A}(\mathbf{r}+\mathbf{r}') S_{X,A}(\mathbf{r'}) 
\phantom{\sum_i}
\right.
\right.
\nonumber \\
& &
{} + \sum_i H(r_i-R')  
\sum_j H(r_j-R') \cdot {}
\nonumber \\
& &
\left.
\left.
\qquad
w_i w_j \phi(\mathbf{r}+\mathbf{r}'-\mathbf{r}_i,\tau_i)
\phi(\mathbf{r}'-\mathbf{r}_j,\tau_j)
\phantom{\sum_i \hspace{-0.2in}}
\right] d^2 r'
\right\rangle.
\end{eqnarray}
Here we have defined $H(x)$ as the Heaviside step function, which is unity for $x>0$ and zero for $x<0$. We have also omitted the limit as $R'\rightarrow \infty$ for brevity; from this point forward we shall understand that all terms are to be evaluated in the limits $R\rightarrow \infty$ and $R'\rightarrow \infty$, with $R \gg R'$. The problem therefore reduces to demonstrating that, in this limit, we have
\begin{eqnarray}
\lefteqn{
\left\langle 
\frac{1}{A}\sum_i H(r_i-R')  
\sum_j H(r_j-R') \cdot {}
\right.
}
\nonumber \\
& &
\left.
\int_A w_i w_j \phi(\mathbf{r}+\mathbf{r}'-\mathbf{r}_i,\tau_i)
\phi(\mathbf{r}'-\mathbf{r}_j,\tau_j)
\, d^2 r'
\phantom{\sum_{i,j}^{r_i>R'\lor r_j>R'}\hspace{-0.65in}}
\right\rangle = 0.
\label{eq:sum_lim}
\end{eqnarray}

Let us first consider the case where event $j$ is outside $A$. Since in this case $|\mathbf{r}'-\mathbf{r}_j| \geq r_j - R'$ for $\mathbf{r}'$ inside $A$, and $\tau_j$ is bounded between 0 and $\tau_f$, we can set an upper limit on the value of $\phi$ inside $A$,
\begin{equation}
\phi(\mathbf{r}'-\mathbf{r}_j,\tau_j) \leq 
\frac{1}{4\pi \tau_0} e^{-(r_j-R')^2/4(\tau_0+\tau_f)}\quad \forall \quad \mathbf{r'} \in A.
\end{equation}
Inserting this upper limit into \autoref{eq:sum_lim}, and noting that the integral of the remaining $\phi$ term over $A$ is bounded above by unity regardless of its arguments (since the integral over all space is always unity), we have
\begin{eqnarray}
\lefteqn{
\int_A w_i w_j \phi(\mathbf{r}+\mathbf{r}'-\mathbf{r}_i,\tau_i)
\phi(\mathbf{r}'-\mathbf{r}_j,\tau_j)
\, d^2 r'
\leq
{}
}
\nonumber \\
& &
\frac{w_i w_j}{4\pi \tau_0} e^{-(r_j-R')^2/4(\tau_0+\tau_f)}.
\end{eqnarray}
If event $i$ is instead the one outside $A$, we can use the same argument to derive a nearly-identical upper limit with $R'$ replaced by $R' - r$. Since we are interested in the limit where $R' \gg r$, this is essentially the same. Thus we have shown
\begin{eqnarray}
\lefteqn{
\left\langle 
\frac{1}{A}\sum_i H(r_i-R')  
\sum_j H(r_j-R') \cdot {}
\right.
}
\nonumber \\
& &
\left.
\int_A w_i w_j \phi(\mathbf{r}+\mathbf{r}'-\mathbf{r}_i,\tau_i)
\phi(\mathbf{r}'-\mathbf{r}_j,\tau_j)
\, d^2 r'
\phantom{\sum_{i,j}^{r_i>R'\lor r_j>R'}\hspace{-0.65in}}
\right\rangle
\nonumber \\
& \leq &
\left\langle 
\frac{1}{A}\sum_i H(r_i-R')  
\sum_j H(r_j-R') \cdot {}
\right.
\nonumber \\
& &
\left.
\frac{w_i w_j}{4\pi \tau_0} e^{-(\max(r_i,r_j)-R')^2/4(\tau_0+\tau_f)}
\phantom{\sum_i\hspace{-0.2in}}
\right\rangle.
\label{eq:sum_lim1}
\end{eqnarray}

Since the expected number of events outside $R'$ approaches $\pi R^2$ for $R \gg R' \gg 1$, and $\max(r_i,r_j) > r$, the expectation value on the right hand side of \autoref{eq:sum_lim1} is limited above by
\begin{eqnarray}
\lefteqn{
\left\langle 
\frac{1}{A}\sum_i H(r_i-R')  
\sum_j H(r_j-R') \cdot {}
\right.
}
\nonumber \\
& &
\left.
\frac{w_i w_j}{4\pi \tau_0} e^{-(\max(r_i,r_j)-R')^2/4(\tau_0+\tau_f)}
\phantom{\sum_i\hspace{-0.2in}}
\right\rangle
\nonumber \\
& \leq &
\frac{\left\langle w_i w_j\right\rangle}{4\tau_0}\left(\frac{R^2}{A}\right)
\int_{R'}^{\infty} p_r(r) e^{-(r-R')^2/4(\tau_0+\tau_f)} \, dr.
\end{eqnarray}
The integral can be evaluated analytically, and in the limit $R \gg R' \gg 1$, the result is
\begin{equation}
\frac{R^2}{A} 
\int_{R'}^{\infty} p_r(r) e^{-(r-R')^2/4(\tau_0+\tau_f)} \, dr
= \frac{2}{R'} \sqrt{\frac{\tau_0+\tau_f}{\pi}}.
\end{equation}
This clearly approaches 0 as $R' \rightarrow \infty$, which demonstrates the required result.

\end{appendix}

\end{document}